\documentclass{gGAF2e}

\usepackage{framed}

\begin{document}




\title{Velocity statistics for point vortices in the local $\alpha$-models of turbulence}

\author{
G. CONTI${\dag}$ $^{\ast}$
and 
G. BADIN${\dag}$ 
\thanks{$^\ast$Corresponding author. Email: giovanni.conti@uni-hamburg.de \vspace{6pt}} 
\\\vspace{6pt}  
${\dag}$ Institute of Oceanography, Center for Earth System Research and Sustainability (CEN), University of Hamburg, Hamburg, Germany
}

\maketitle

\begin{abstract}
The velocity fluctuations for point vortex models are studied for the $\alpha$-turbulence equations, which are characterized by a fractional Laplacian relation between active scalar and the streamfunction. In particular, we focus on the local dynamics regime.
The local dynamics differ from the well-studied case of $2$D turbulence as it allows to consider the true thermodynamic limit, that is, to  consider an infinite set of point vortices on an infinite plane keeping the density of the vortices constant. The consequence of this limit is that the results obtained are independent on the number of point vortices in the system. This limit is not defined for $2$D turbulence. 
We show an analytical form of the probability density distribution of the velocity fluctuations for different degrees of locality. 
The central region of the distribution is not Gaussian, in contrast to the case of $2D$ turbulence, but can be approximated with a Gaussian function in the small velocity limit.
The tails of the distribution exhibit a power law behavior and  self similarity in terms of the density variable. 
Due to the thermodynamic limit, both  the Gaussian approximation for the core and the steepness of the tails are independent on the number of point vortices, but depend on the $\alpha$-model.   
We also show the connection between  the velocity  statistics for point vortices uniformly distributed, in the context of the $\alpha$-model in classical turbulence, with the velocity statistics  for point vortices non-uniformly distributed. 
Since the exponent of the power law depends just on $\alpha$, we test the power law approximation obtained with the point vortex approximation, by simulating  full turbulent fields for different values of $\alpha$ and we compute the correspondent probability density distribution for the absolute value of the velocity field.
 These results suggest that the local nature of the turbulent fluctuations in the ocean or in the atmosphere might be deduced from the shape of the tails of the probability density functions.
\end{abstract}


\section{\label{sec:level1}Introduction} 
Turbulent flows, such as the one observed in astrophysical, geophysical, pipe as well as quantum flows, are characterized by inertial ranges which extend through several scales of motion. In the atmosphere and the ocean, for example, the quasi two-dimensional dynamics has inertial ranges that extend from the large scales, to synoptic and until to submesoscale (e.g. \citet{charney71}). Although  the development and representation of turbulence is largely unknown, this physical phenomenon drives many important processes such as the transfer of energy to the dissipative scales and the stirring and mixing of tracers. It is then of primary importance trying to understand further the features and the representation of  turbulence. In particular, the statistical representation of velocity fluctuations may be used for the formulation of stochastic models representing the dynamics. However, trying to solve analytically these problems is not simple, and simplifications are often required. 

Depending on the model used to study turbulence, inverse or forward cascades of the generalized energy and enstrophy can occur in the turbulent inertial ranges. This transfer is manifested with the creation of vorticity dominated structures. For this reason the vorticity dynamics plays a crucial role in the description of turbulence. If we limit our investigation to $2$D flows, 
we can introduce an important simplification through the use of point vortices. In this approximation the vorticity field is approximated by a set of localized point vortices, which is the analogous of replacing a continuous mass distribution by a set of localized material points. The resulting dynamics is Hamiltonian and is characterized by the conservation of energy, as well as the linear and angular momentum \citep{helmholtz1858}.  See e.g. \citet{aref2007point,marchioro2012mathematical,newton2013n} for reviews and \citet{chapman1978ideal,badin2018variational} for a discussion on symmetries and conservation laws. For analogies with electrodynamics, see e.g. \citet{Kraichnan1980}, for applications to astrophysical problems see e.g. \citet{Chandra1941,Chandra1942,Chandra1943,Chavanis1998Jupiter}, for the statistical dynamics of point vortices in 2D fluid dynamics see e.g. \citet{Onsager1949,Weiss1998,Chavanis1999,buhler2002statistical,esler2017equilibrium}.  

The interactions between point vortices is determined by the model used to study the dynamics.
In this work we investigate the velocity field produced by a family of models, the $\alpha$-models. These models are defined using a generalized Euler equation
\begin{equation}
\frac{\partial\zeta}{\partial t}+J(\psi,\zeta)=0, 
\label{geul}
\end{equation}
where $\zeta(x,y,t)$ is an active scalar, $\psi(x,y,t)$ is the streamfunction, and $J(\psi,\zeta)=\partial_x \psi \partial_y \zeta - \partial_x \zeta \partial_x \psi$ is the Jacobian determinant. The form of the coupling between these two fields determines the degree of locality of the model. The streamfunction and active scalar fields are related by
\begin{equation}
\zeta=-(-\bm{\Delta}\psi)^{\alpha/2},
\label{coupl}
\end{equation}
where $\bm{\Delta}$ is the $2$D Laplacian operator. In the Fourier space \eqref{coupl} becomes
\begin{equation}
\hat{\psi}(\bm{k})=-\lvert\bm{k}\rvert^{-\alpha}\hat{\zeta}(\bm{k}),
\end{equation}
where $\bm{k}$ is the $2$D wavenumber vector, with magnitude $k=\lvert\bm{k}\rvert$. When $\alpha$ increases, the fields become more decoupled and the problem becomes more spectrally nonlocal. The case $\alpha=2$ represents the widely studied Euler equation, while $\alpha<2$ and $\alpha>2$ represent respectively the local and nonlocal dynamics. For studies of different forms of turbulence emerging from different values of $\alpha$  see, e.g.  \citet{pierrehumbertetal94,smith2002turbulent,tran2002constraints,tran2004nonlinear,tran2010effective,burgess2013spectral,burgess2015kraichnan,Norbert2000,venaille2015violent,foussard2017relative,BadinBarry2018}. 

The special case of $\alpha=1$ can be used to study the local  dynamics of a stratified rapidly rotating flow with zero potential vorticity in the interior domain, and assigned potential temperature at the surface (e.g. the atmospheric tropopause or the oceanic surface). In this case the scalar field $\zeta$ represents the potential temperature. This model is called Surface Quasi Geostrophic (SQG) model \citep{blumen78,heldetal95,lapeyre17,badin2018variational}. The analysis of the kinetic energy spectra for this model shows a characteristic forward energy cascade \citep{tulloch2006theory,capet2008surface}, with consequent formation of small structures in the flow. This behaviour is complementary to the one of the $2$D turbulence, and makes the SQG model a possible candidate for the forward cascade of temperature variance and the formation of frontal structures. SQG is also a candidate for the explanation of the submesoscale dynamics (where the term submesoscale is here used in its oceanographic meaning, see e.g. \citet{McWilliams2016}), which is also important for the mixing of passive tracers \citep{badin2011lateral,Shcherbina2015,Mukiibi2016}.  For a study on the relationship between quasi geostrophic (QG) and SQG turbulence, see e.g. \citet{badin2014role}. For the Hamiltonian and Nambu structure of SQG, see e.g. \citet{blender2015hydrodynamic}. The stability of SQG vortices was studied by e.g. \citet{carton2009instability,dritschel2011exact,harvey2011perturbed,harvey2011instability,bembenek2015realizing,carton2016vortex,Badinpoulin2018}. SQG point vortices have been studied by e.g. \citet{lim2001point,taylor2016dynamics,BadinBarry2018}.

Mathematically, SQG shows interesting analogies with the $3$D Euler equation \citep{constantinetal94}. This analogy sparked a large interest, as it suggests that the study of the regularity of the SQG model could provide hints for the formation of singularities in the 3D Euler equation, see e.g. \citet{constantinetal94,constantin1994singular,pierrehumbertetal94,majda1996two,ohkitani1997inviscid,constantin1998nonsingular,constantin1999behavior,cordoba2002growth,cordoba2002scalars,cordoba2004maximum,rodrigo2004vortex,cordoba2005evidence,rodrigo2005evolution,wu2005solutions,deng2006level,dong2008finite,ju2006geometric,li2009existence,marchand2008existence,marchand2008weak,scott2011scenario,constantin2012new,ohkitani2012asymptotics,scott2014numerical,BadinBarry2018}.  

The aim of this work is the investigation of velocity statistics for turbulent flow, with the simplification of point vortices randomly distributed and an interaction between vortices ruled by a family of $\alpha$-model. One important difference that emerges from the computation is that, away from $\alpha=2$, it is really possible to consider a proper thermodynamics limit, that is,  we can consider an infinite set of point vortices on an infinite plane keeping the density of the vortices constant. As a consequence of this limit, all the results we obtain are independent on the number  of point vortices in the domain considered.
Given their physical interest, we will focus on local  dynamics, with particular attention to the SQG model. 
Throughout the article we will follow closely the approach used by \citet{Chavanis1999,Skaugen2016} to study $2$D turbulence using a random distribution of point vortices, and the one of \citet{Chavanis2009} to study the statistics of the gravitational force induced by inhomogeneous distribution of sources in $d$ dimensions. To test the reliability of the point vortex approximation, we have also simulated \eqref{geul} and \eqref{coupl} for three different values of $\alpha$ computing then the corresponding probability density functions (PDFs) for the absolute value of the velocity fields to compare with their analytical approximation.

\section{\label{sec:level2}Statistical Formulation}

\subsection{\label{sec:levelformal} Formal solution}

Consider a point vortex with circulation $\gamma$ placed at $\bm{r}_0$, so that
\begin{equation}
\zeta(\bm{r})=\gamma\, \delta(\bm{r}-\bm{r}_0),
\label{vor}
\end{equation}
where $\delta$ is the Dirac delta function and $\zeta$ the active scalar field at the position $\bm{r}$. Even if $\zeta$ can represent active scalars which might have a different physical meaning than vorticity, throughout the article we will still call the objects arising from relation \eqref{vor} as point "vortices".

For the $\alpha$-models, the Green's functions $G^{(\alpha)}(\bm{r})$ are  introduced following \citet{Iwayama2010} as
\begin{align}
G^{(2)} \left( \bm{r},\bm{r}' \right)&=\frac{1}{2\pi}\ln \left(\bm{r}-\bm{r}' \right) + C~,\label{g1}\\
G^{(\alpha)} \left( \bm{r},\bm{r}' \right)&=\Psi \left(\alpha \right) \left( \bm{r}-\bm{r}' \right)^{\alpha-2}~, \qquad \alpha\neq2~,\label{g2}
\end{align}
with $C$  arbitrary constant and 
\begin{equation}
\Psi \left( \alpha \right)=-\left\{ 2^{\alpha}\left[\Gamma \left(\frac{\alpha}{2} \right)\right]^2 \sin\left(\frac{\alpha \pi}{2}\right) \right\}^{-1}~,
\label{psi}
\end{equation}
so that the streamfunction of the flow can be expressed as
\begin{equation}
\psi \left(\bm{r} \right)=\int G^{\left(\alpha \right)} \left(\bm{r},\bm{r}' \right)\zeta \left(\bm{r}' \right) \,d \bm{r}'\,.
\end{equation}
 
\begin{figure}
\centering
\includegraphics[width=.5\textwidth]{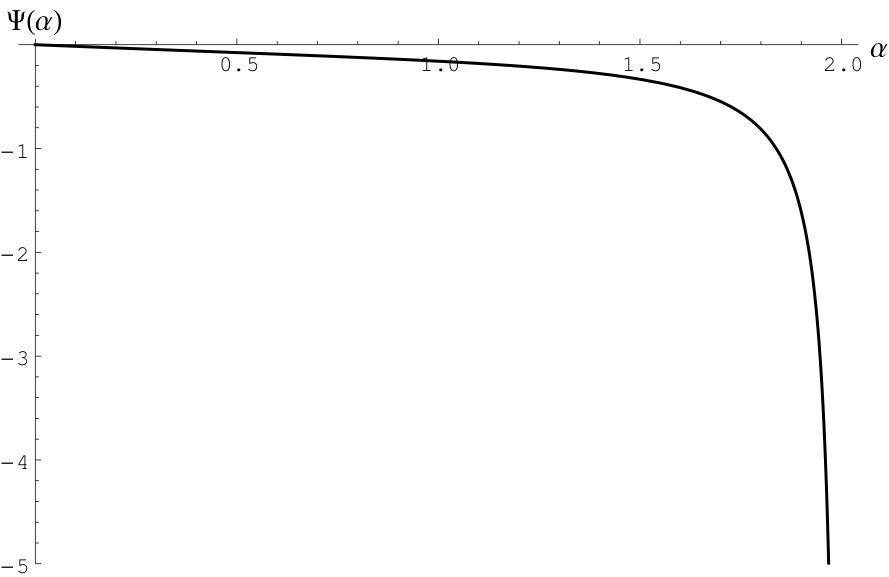}
\caption{$\Psi \left(\alpha \right)$, as defined by \eqref{psi}, for different $\alpha$-models with $\alpha\in \left(0,\,\,2 \right)$. }
\label{fig:psi}
\end{figure}

Figure \ref{fig:psi} shows the behaviour of $\Psi(\alpha)$, for different values of $\alpha$. In the considered interval, $\Psi(\alpha)$ results to be always negative. Notice that when $\alpha\to2$ then $\Psi(\alpha)\to-\infty$,  there is a discontinuity in the Green's functions between the local, $\alpha<2$, and nonlocal, $\alpha>2$, dynamics.
Using the stream function we can write the velocity field generated in a given location by the point vortex following \citet{BadinBarry2018} as
\begin{align}
\bm{\phi}^{(2)}&=-\frac{\gamma}{2\pi}\frac{\bm{r}_\perp}{r^2}\label{vr0}\\
\bm{\phi}^{(\alpha)}&=-\gamma\Psi(\alpha)\frac{\bm{r}_\perp}{r^{(4-\alpha)}}~, \qquad \alpha\neq2~.
\label{vr}
\end{align}
In \eqref{vr}, the $\perp$ subscript  denotes the clockwise rotation of a vector by $\pi/2$. 

In the following we will consider  an ensemble of $N$ identical point vortices, and following \citet{Chavanis1999,Skaugen2016} we shall assume:
\begin{itemize}
 \item[(i)] a ``\emph{neutral}'' system, i.e. a system in which its vortices can exhibit only two values for the circulation, $+\lvert\gamma\lvert$ and $-\lvert\gamma\lvert$, in equal proportion to avoid solid rotation;
 \item[(ii)] vortices randomly distributed with uniform probability  on a disk of radius $R$;
 \item[(iii)]   uncorrelated vortices, that is, the positions of the vortices inside the domain are uncorrelated and the probability of the N-point configurational distribution can be seen as a product of the probability $p(\bm{r})$ of finding a vortex.
\end{itemize}

Assumptions (i) and (ii) state that a vortex with circulation $-\lvert\gamma\lvert$ located in  $\bm{r}$ produces the same velocity field of a vortex with circulation $+\lvert\gamma\lvert$ located in $-\bm{r}$, and since the two species of vortices are randomly distributed with uniform probability we have the \textit{statistical equivalence of the two groups of vortices}, and we can proceed further considering just a single species of vortices. 
Assumption (iii) is used in defining the probability density for the velocity, as we will show in a moment.

We shall also assume:
\begin{itemize}
\item[(iv)] that, without loss of generality, the velocity $\bm{V}$ is computed at the center of the domain.
\end{itemize}

From (ii)  follows that the probability density of having a vortex at position $\bm{r}$ is
\begin{equation}
 p(\bm{r})=\frac{1}{\pi R^2},
 \label{unipdf}
\end{equation}
and the vortices density is
\begin{equation}
n=\frac{N}{\pi R^2}.
\label{dens}
\end{equation}

The velocity field $\bm{V}$ produced by the $N$ vortices at a given point is the sum of the velocity fields at that point 
\begin{equation}
\bm{V}=\sum_{i=1}^{N} \bm{\phi}_i^{(\alpha)}, \qquad \alpha \in (0,\,\,2].	
\end{equation}

Under these considerations the PDF $P_N^{(\alpha)}(\bm{V})$, that gives the probability $P_N^{(\alpha)}(\bm{V})d\bm{V}$ for the velocity $\bm{V}$ to fall between $\bm{V}$ and $\bm{V}+d\bm{V}$, 
can be written as
\begin{equation}
P_N^{(\alpha)}(\bm{V})=\int \prod_{i=1}^{N} p(\bm{r}_i)\, \delta\left(\bm{V}^{}-\sum_{i=1}^{N} \bm{\phi}_i^{(\alpha)}\right)\,d\bm{r},
\label{pdf}
\end{equation}
where the assumption (iii) has been used to factorized the PDF for the N-point configuration.
To simplify \eqref{pdf}, decoupling the integral, we can use the integral form of the Dirac delta function
\begin{equation}
\delta(\bm{x})=\frac{1}{(2\pi)^2}\int e^{-i \bm{\rho} \cdot \bm{x}}\,d\bm{\rho}.
\end{equation}
Inserting this expression in \eqref{pdf} yields
\begin{equation}
P_N^{(\alpha)}(\bm{V}^{})=\frac{1}{4\pi^2}\int A_N^{(\alpha)}(\bm{\rho}) \,e^{-i \bm{\rho}\cdot\bm{V}^{}}\,d\bm{\rho}
\end{equation}
with
\begin{align}
A_N^{(\alpha)}(\bm{\rho})&=\left(\int_{\lvert\bm{r}\rvert=0}^R e^{i\bm{\rho\cdot\bm{\phi}^{(\alpha)}}}p(\bm{r}) \,d\bm{r}\right)^N \nonumber \\
&=\left(\frac{1}{\pi R^2}\int_{\lvert\bm{r}\rvert=0}^R e^{i\bm{\rho\cdot\bm{\phi}^{(\alpha)}}} \,d\bm{r}\right)^N
\label{AN}
\end{align}
where we have made use of \eqref{unipdf}.  Since the probability density for the vortices position is normalized to the unity, \eqref{AN} can
be written as
\begin{align}
A_N^{(\alpha)}(\bm{\rho})&=\left(1-\frac{1}{\pi R^2}\int_{\lvert\bm{r}\rvert=0}^R \left(1-e^{i\bm{\rho\cdot\bm{\phi}^{(\alpha)}}} \right) \,d\bm{r}\right)^N \nonumber \\
&=\left(1-\frac{n}{N}\int_{\lvert\bm{r}\rvert=0}^R \left(1-e^{i\bm{\rho\cdot\bm{\phi}^{(\alpha)}}} \right) \,d\bm{r}\right)^N .
\label{ANL}
\end{align}
When the thermodynamic limit is considered, that is
\begin{equation}
  N\to\infty, \qquad R\to\infty, \quad \text{with}\quad n=const,
  \label{tlim}
 \end{equation} 
from the special limit in \eqref{ANL} we obtain
\begin{equation}
A^{(\alpha)}(\bm{\rho})=e^{-n\, C^{(\alpha)}(\bm{\rho})},
\label{alim}
\end{equation}
with
\begin{equation}
C^{(\alpha)}(\bm{\rho})=\int_{\lvert\bm{r}\rvert=0}^\infty \left(1-e^{i\bm{\rho}\cdot\bm{\phi}^{(\alpha)}} \right)\,d\bm{r}~,
\label{C}
\end{equation}
where we have dropped the subscript N to indicate that the thermodynamics limit was been taken.

Notice that for $\alpha=2$ the thermodynamic limit \eqref{tlim} does not exist due to logarithmic divergence with the number of vortices \citep{jimenez_1996,Min1996,Weiss1998}. Therefore, in the particular case of $2$D turbulence, \eqref{alim} must not be regarded as a true limit. For the same reason, for $\alpha = 2$ it is not possible to consider the upper limit in the integral \eqref{C} as equal to infinity.

In order to proceed further in the computation of $P^{(\alpha)}(\bm{V}^{})$, we have to evaluate $C^{(\alpha)}$. As suggested by \citet{Chavanis1999,Chavanis2009} it is convenient to carry on the computation changing the variable of integration from $\bm{r}$ to $\bm{\phi}$. From  \eqref{vr0} and \eqref{vr} it follows that the Jacobi determinant of the transformation is
\begin{align}
\left\lVert \frac{\partial(\bm{r})}{\partial \left(\bm{\phi}^{(2)} \right)}\right\rVert &= -\frac{\gamma^2}{4\pi^2}\left( \phi^{(2)} \right)^{-4}~,\\
\left\lVert \frac{\partial(\bm{r})}{\partial \left(\bm{\phi}^{(\alpha)} \right)}\right\rVert &= -\frac{\left(\gamma \lvert\Psi(\alpha)\rvert \right)^\frac{2}{3-\alpha}}{3-\alpha}\left( \phi^{(\alpha)} \right)^{-2\left(\frac{4-\alpha}{3-\alpha}\right)}~,\qquad \alpha\neq 2~,
\end{align}
and the integrals \eqref{C} become
\begin{equation}
C^{(2)}(\bm{\rho})=\frac{\gamma^2}{4\pi^2}\int_{\lvert\bm{\phi^{(2)}}\rvert=\frac{\gamma}{2\pi R}}^{\infty}\left( 1-e^{i\bm{\rho}\cdot\bm{\phi}^{(2)}} \right) \left(\phi^{(2)} \right)^{-4}\,d\bm{\phi}^{(2)}~,
\label{C2start}
\end{equation}
and
\begin{align}
C^{(\alpha)}(\bm{\rho})=
\frac{\left(\gamma \lvert\Psi(\alpha)\rvert \right)^\frac{2}{3-\alpha}}{3-\alpha} 
 \int_{\lvert\bm{\phi^{(\alpha)}}\rvert=0}^{\infty}\left(1-e^{i\bm{\rho}\cdot\bm{\phi^{(\alpha)}}} \right) \left(\phi^{(\alpha)} \right)^{-2\left(\frac{4-\alpha}{3-\alpha}\right)}\,d\bm{\phi}^{(\alpha)}~, ~
\alpha\neq 2~.
 \label{Ca1}
\end{align}
We first consider \eqref{C2start}. Switching to polar coordinates, choosing the radial coordinate  in the direction of $\bm{\rho}$, and defining with $\theta$ the angle between $\bm{\rho}$ and $\bm{\phi}^{(2)}$, we obtain
\begin{align}
C^{(2)}(\bm{\rho})&=\frac{\gamma^2}{(2\pi)^2}\int_{\frac{\gamma}{2\pi R}}^{\infty} \int_0^{2\pi} \left(1-e^{i\rho\phi\cos\theta} \right)\left(\phi^{(2)} \right)^{-3}\,d\theta\,d\phi^{(2)} \nonumber \\
                            &=\frac{\gamma^2}{2\pi}\int_{\frac{\gamma}{2\pi R}}^{\infty}  \left(1-J_0 \left(\rho\phi^{(2)} \right) \right)\left(\phi^{(2)} \right)^{-3}\,d\phi^{(2)},
\end{align}
where $J_0$ is the Bessel function of the first kind and order zero. 
Finally, substituting $x=\rho\phi^{(2)}$,
\begin{align}
C^{(2)}(\bm{\rho})&=2\pi\left(\frac{\gamma\rho}{2\pi}\right)^2\int_{\frac{\gamma}{2\pi R}}^{\infty} \left(1-J_0(x) \right) x^{-3}\,dx \nonumber \\
                            &=2\pi\, \left(\frac{\gamma\rho}{2\pi}\right)^2 \, \kappa^{(2)}(R) ,
\label{C2}
\end{align}
where $\kappa^{(2)}(R)$ is a dimensionless  number given by the integral in \eqref{C2}. 
Using a similar strategy for the case $\alpha\neq2$,  we switch to polar coordinate and extend the limits of the integration in \eqref{Ca1} to get
\begin{align}
 C^{(\alpha)}(\bm{\rho})&=\frac{\left(\gamma \lvert\Psi(\alpha)\rvert \right)^\frac{2}{3-\alpha}}{3-\alpha} 
                           \int_{0}^{\infty} \int_0^{2\pi} 
                         \left(1-e^{i\rho\phi\cos\theta} \right)
                         \left(\phi^{(\alpha)} \right)^{-\left(\frac{5-\alpha}{3-\alpha}\right)}\,d\theta\,d\phi^{(\alpha)}\nonumber \\
                            &=2\pi\frac{\left(\gamma \lvert\Psi(\alpha)\rvert \right)^\frac{2}{3-\alpha}}{3-\alpha}
         \int_{0}^{\infty}  \left(1-J_0 \left(\rho\phi^{(\alpha)}\right) \right) \left(\phi^{(\alpha)} \right)^{-\left(\frac{5-\alpha}{3-\alpha}\right)} \,d\phi^{(\alpha)}, ~
         \alpha\neq2  .
\end{align}
Substituting $x=\rho\phi^{(\alpha)}$ yields
\begin{align}
C^{(\alpha)}(\bm{\rho})&=2\pi A^{(\alpha)} \rho^{\frac{2}{3-\alpha}}
         \int_{0}^{\infty} \left(1-J_0(x) \right) x^{-\left(\frac{5-\alpha}{3-\alpha}\right)} \,dx \nonumber \\
                            &=2\pi \,A^{(\alpha)} \, \kappa^{(\alpha)} \, \rho^{\frac{2}{3-\alpha}}, \quad\alpha\neq2,
\label{Ca}
\end{align}
where 
\begin{equation}
A^{(\alpha)}=\frac{(\gamma \lvert\Psi(\alpha)\rvert)^\frac{2}{3-\alpha}}{3-\alpha}~,
\end{equation}
and $\kappa^{(\alpha)}$ is a dimensionless  number given by the integral in \eqref{Ca}, that depends parametrically on $\alpha$.

The formal solution for the PDF can thus be written as
\begin{align}
P^{(\alpha)}(\bm{V})&=\frac{1}{(2\pi)^2}\int A^{(\alpha)}(\bm{\rho})e^{i\bm{\rho}\cdot\bm{V}}\,d\bm{\rho} \nonumber \\
                                &=\frac{1}{(2\pi)^2}\int_0^{2\pi}\int_0^{\infty}\rho e^{-n C^{(\alpha)}(\bm{\rho})}e^{i\rho V\cos\theta}\,d\rho\, d\theta~,
\label{formalsol}
\end{align}
with $\alpha \in (0,\,\,2]$, and $C^{(\alpha)}$ given by either \eqref{C2} or \eqref{Ca}, and the introduction of polar coordinates. 
From now on we concentrate on the interval $\alpha\in(0,\,2)$ if not explicitly specified differently.

The function $\kappa^{(\alpha)}$ that appears in \eqref{Ca} can be evaluated as (see, e.g. \citet{Skaugen2016})
\begin{align}
&\kappa^{(\alpha)}=\int_0^{\infty} (1-J_0(x)) x^{-\left(\frac{5-\alpha}{3-\alpha}\right)} \,dx \nonumber \\
           &=-\frac{1}{\pi}\sin\left[\frac{\pi}{2}\left(\frac{1-\alpha}{3-\alpha}\right)\right]\Gamma\left(\frac{-2}{3-\alpha}\right)B\left[\frac{1}{2}\left(\frac{5-\alpha}{3-\alpha}\right),\frac{1}{2}\right],
\label{kappa}
\end{align}
where $B$ is the Beta function. 

\begin{figure}
\centering
\includegraphics[width=.5\textwidth]{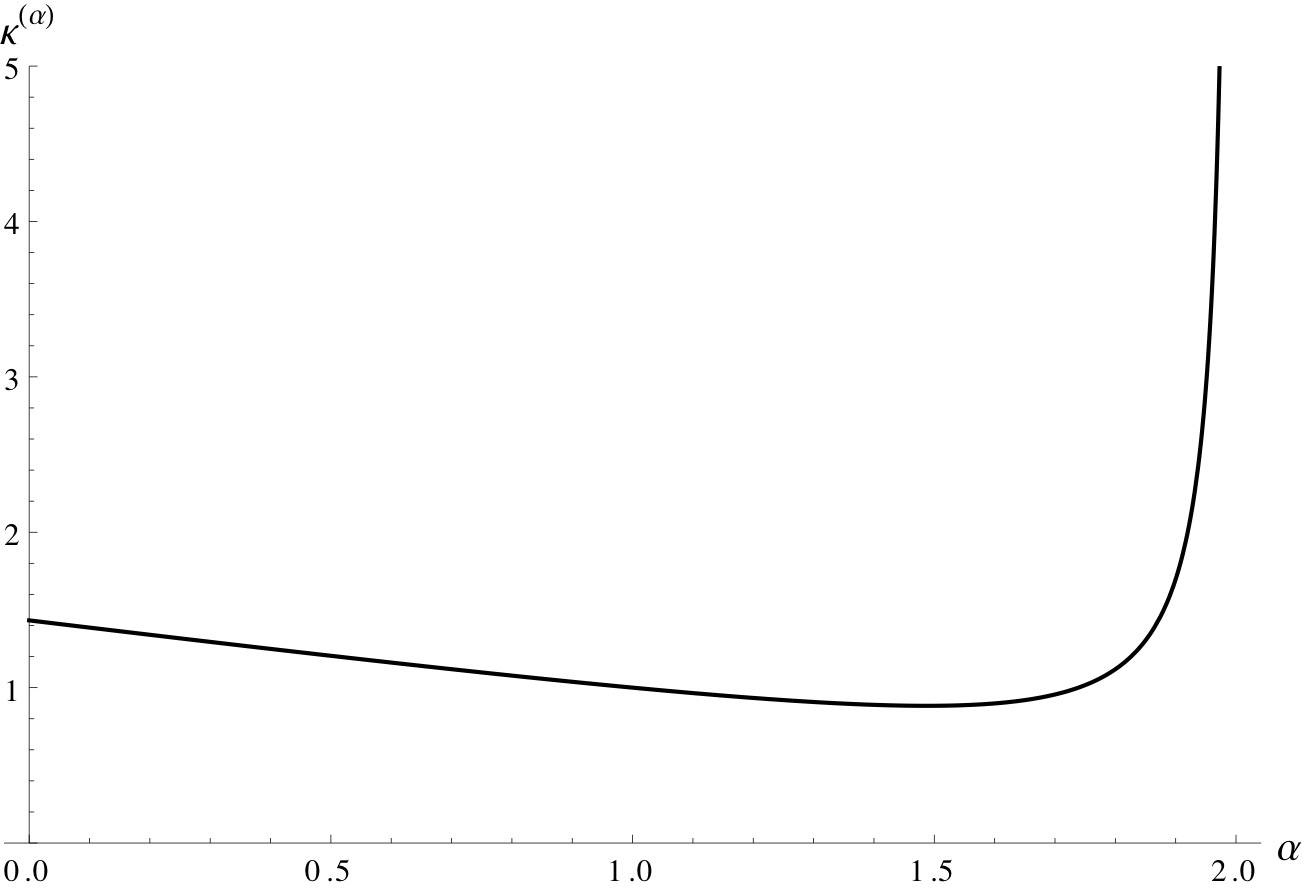}
\caption{$\kappa^{(\alpha)}$, as defined by \eqref{kappa}, for different $\alpha$-models with $\alpha\in \left(0,\,\,2 \right)$.}
\label{fig:kappa}
\end{figure}

Figure \ref{fig:kappa} shows the behaviour of $\kappa^{(\alpha)}$ for different $\alpha$-models with $\alpha\in(0,\,\,2)$. The function $\kappa^{(\alpha)}$ shows positive values in all the interval, decreasing for increasing values of $\alpha$ until  $\alpha\approx1.48$, where it exhibits a minimum, and then increasing for the rest of the interval. The function diverges for $\alpha\to2$. 

Because $\kappa^{(\alpha)}>0$  in all the $(0,\,\,2)$ interval, for these values of $\alpha$ we can solve for the tails of the distribution following the same procedure used by \citet{Chavanis1999,Skaugen2016}. This will be done in the next Section.


\subsection{\label{sec:leveltail} Tails of the distribution}

In order to study the tails of the distribution, we start by considering the following physical argument. High velocities in the tails are induced only by nearby vortices. As the probability to find nearby vortices to a given vortex is low, in first approximation only one neighbouring vortex will be important. 
The probability that the nearest neighbour occur between $r$ and $r+dr$ is $p^{(\alpha)}(r)dr$, where
\begin{equation}
p^{(\alpha)}(r) = 2\pi n r e^{-\pi n r^2}.
\label{ndistfull}
\end{equation}
In \eqref{ndistfull}, the superscript $\alpha$ indicates the implicit dependence on the degree of locality of the model in the inversion between the distance between vortices $r$ and the velocity $V$.

When $r$ is small, the nearest neighbour distribution becomes
\begin{equation}
p^{(\alpha)}_0(r) \sim 2\pi n r.
\label{ndist}
\end{equation}
The subscript $0$ indicates that \eqref{ndist} is the leading term in the expansion for small $r$ of the full nearest neighbour distribution. With these premises we can write thus for the tails
\begin{equation}
P^{(\alpha)}(\lvert\textbf{V}\rvert=V)\sim p_0^{(\alpha)}(r(V))\left\lvert\frac{d r(V)}{dV}\right\lvert
\label{changevarpdf}
\end{equation}
that is
\begin{equation}
P^{(\alpha)}(\lvert\bm{V}\lvert=V)\sim\frac{2\pi n \left(\gamma\lvert\Psi(\alpha)\rvert\right)^{\frac{2}{3-\alpha}}}{3-\alpha} V^{-\frac{5-\alpha}{3-\alpha}}.
\label{tailmod}
\end{equation}
From \eqref{tailmod} it is visible that the absolute value of the exponent of the power law tail increases thus with $\alpha$. 

The PDF for the absolute value of the velocity for a fixed value of $\lvert\bm{V}\lvert=V$ is related to the PDF of the velocity by
 $P^{(\alpha)}(\lvert\bm{V}\lvert=V)= 2\pi V P^{(\alpha)}(\bm{V})$, 
so that
\begin{equation}
P^{(\alpha)}(\textbf{V})\sim  \frac{n \left(\gamma\lvert\Psi(\alpha)\rvert\right)^{\frac{2}{3-\alpha}}} {3-\alpha}V^{-2\frac{4-\alpha}{3-\alpha}}.
\label{tail}
\end{equation}


The distribution \eqref{ndist} can be recovered from \eqref{ndistfull} not only when $r$ is small, but also when $\pi n r^2$ is small. This implies that approximation \eqref{ndist} is valid also when the distance of the nearest neighbour is not small but the vortices are diluted in the domain. The magnitude of the velocity for which \eqref{tail} start to fail can be calculated expanding the exponential in \eqref{ndistfull} to the first order and determining when the first order approximation has half of the weight of the leading order term. Then
\begin{equation}
p^{(\alpha)}(r) \sim p_0^{(\alpha)}(r) +  p_1^{(\alpha)}(r)= 2\pi n r - 2 \pi^2 n^2 r^3 ~, 
\end{equation}
and, using \eqref{changevarpdf}, 
\begin{equation}
P^{(\alpha)}(\lvert\bm{V}\lvert=V)\sim P_0^{(\alpha)}(\lvert\bm{V}\lvert=V) - P_1^{(\alpha)}(\lvert\bm{V}\lvert=V)~,
\end{equation}
where $P_0^{(\alpha)}$ is the right hand side of \eqref{tailmod} and
\begin{equation}
P_1^{(\alpha)}(\lvert\bm{V}\lvert=V) = \frac{2\pi^2 n^2 \left(\gamma\lvert\Psi(\alpha)\rvert\right)^{\frac{4}{3-\alpha}}}{3-\alpha} V^{-\frac{7-\alpha}{3-\alpha}}~.
\end{equation}
Using the condition 
\begin{equation}
P_1^{(\alpha)}(\lvert\bm{V}\lvert=V_{cut})=\frac{1}{2}P_0^{(\alpha)}(\lvert\bm{V}\lvert=V_{cut})~,
\end{equation}
one obtains
\begin{equation}
V_{cut} = (2 n\pi)^{\frac{3-\alpha}{2}} \gamma \lvert\Psi(\alpha)\rvert~.
\label{Vcut}
\end{equation}


For strongly local dynamics one observes the formation of  clusters of vortices (see e.g. the numerical simulations shown in Section \ref{sec:Numerical Simulations}), so that the physical argument used at the beginning of this Section, which is based on the presence of a single nearest neighbour, does not hold any longer. Even in this case, it is however still possible to obtain analytical results which are independent on the assumption of a single nearest neighbour. These results can be obtained from the explicit evaluation of some integrals for the specific case $V\gg1$. It is shown in the Appendix that in this asymptotic regime one recovers the shape for the tails of the PDF given by \eqref{tail}. 
%
%
%

\subsubsection{Self-similarity of the tails}

The results so far obtained show an interesting self-similarity of the tails of the PDF with respect to the density $n$. 
If we rescale the velocity using the density
\begin{equation}
\tilde{\bm{V}}=\bm{V}/n^{\frac{3-\alpha}{5-\alpha}}
\end{equation}
the PDF of the velocity can be rewritten as
\begin{equation}
P^{(\alpha)}(\bm{V})\sim n^{-\frac{3-\alpha}{5-\alpha}}\mathcal{P}\left(\bm{V}/n^{\frac{3-\alpha}{5-\alpha}}\right),
\end{equation}
which shows that the PDF is self-similar with respect to the density $n$. The rescaled distribution is
\begin{equation}
\mathcal{P}(\tilde{\bm{V}})\sim V^{-2\frac{4-\alpha}{3-\alpha}}~,
\end{equation}
which is independent of $n$.

Similarly, for the probability density of the module of the velocity we can set
\begin{equation}
\tilde{\bm{V}}= \bm{V}/n^{\frac{3-\alpha}{2}}
\end{equation}
so that 
\begin{equation}
P^{(\alpha)}(\lvert\bm{V}\lvert=V)\sim n^{-\frac{3-\alpha}{2}}\mathcal{P}\left(\bm{V}/n^{\frac{3-\alpha}{2}}\right),
\end{equation} 
and 
\begin{equation}
\mathcal{P}(\tilde{\bm{V}})\sim \tilde{V}^{-\frac{5-\alpha}{3-\alpha}}~,
\end{equation} 
which is independent of $n$.

This self-similarity is particularly meaningful for the $\alpha$-models thanks to the validity of the thermodynamics limit, following which the results do not depend on the number of vortices but on their density $n$.

\subsection{\label{sec:levelcore} Core of the distribution and small velocity limit}

In order to explore the core of the distribution we can integrate the angular part of \eqref{formalsol} to obtain
\begin{equation}
P^{(\alpha)}(\bm{V})=\frac{1}{2\pi}\int_{0}^{\infty}\exp \left({-c^{(\alpha)}\rho^{\frac{2}{3-\alpha}}} \right)\, \rho \, J_0(\rho V)\,d\rho,
\label{pfull}
\end{equation}
where we have set
\begin{equation}
c^{(\alpha)}=n 2\pi \kappa^{(\alpha)}\frac{(\gamma \lvert\Psi(\alpha)\rvert)^\frac{2}{3-\alpha}}{3-\alpha},
\label{ca}
\end{equation}
and with $\kappa^{(\alpha)}$ defined in \eqref{kappa}.
Since the core of the distribution describes the behaviour for small velocity, we can expand the Bessel function for small argument $\rho V \ll 1$
to get
\begin{align}
P^{(\alpha)}(\bm{V})\sim\frac{1}{2\pi}\int_{0}^{\infty}\exp \left({-c^{(\alpha)}\rho^{\frac{2}{3-\alpha}}} \right)
 \rho \left(1-\frac{(\rho V)^2}{4}+\frac{(\rho V)^4}{64}\right)\,d\rho.
\end{align}
The evaluation of the integral above yields the PDF
\begin{equation}
P^{(\alpha)}(\bm{V})\sim \frac{\left( c^{(\alpha)} \right)^{\alpha-3}\Gamma(4-\alpha)}{4\pi} 
\left(1-\frac{\left( c^{(\alpha)} \right)^{\alpha-3}}{8} \frac{\Gamma(7-2\alpha)}{\Gamma(4-\alpha)}V^2 
+\frac{\left( c^{(\alpha)} \right)^{2(\alpha-3)}}{192} \frac{\Gamma(10-3\alpha)}{\Gamma(4-\alpha)}V^4 \right) .
\label{coreexp}
\end{equation}
The form of the PDF \eqref{coreexp} differs from a Gaussian Taylor series already by the third term in the expansion, implying that the core of the distribution is not Gaussian, differing then from the case  $\alpha=2$.
 
However, in the small velocity limit the distribution can be approximated to a Gaussian function considering 
 only the first  two terms in the expansion \eqref{coreexp} that is
\begin{equation}
P^{(\alpha)}(\bm{V})\sim\frac{\left( c^{(\alpha)} \right)^{\alpha-3}\Gamma(4-\alpha)}{4\pi} 
\exp \left(-\frac{1}{4  \left(c^{(\alpha)} \right)^{3-\alpha}}\frac{\Gamma(6-2\alpha)}{\Gamma(3-\alpha)}V^2\right).
\label{ga2}
\end{equation}
%
%
Notice that when $\alpha\to0$ the core of the distribution becomes more peaked to highlight that the dynamics is more local and the velocity field between the vortices is characterized mostly by lower velocity. 


Similarly to the analysis for the tails of the distribution, we can ask for which value of $V$ the  Gaussian function is no longer a valid approximation. Equating the quartic term in \eqref{coreexp} with half of the quadratic term yields
\begin{equation}
V_{gcut}= 2 \sqrt{3}\left(c^{(\alpha)} \right)^{\frac{3-\alpha}{2}}\sqrt{\frac{\Gamma(7-2\alpha)}{\Gamma(10-3\alpha)}}.
\label{Vgcut}
\end{equation}
As shown in the next Section, $V_{gcut}$ increases with $\alpha$. For $\alpha\to2$ the value of $V_{gcut}$ instead diverges. This divergence is  the consequence of the divergence of the Gaussian core of the distribution for $2D$ turbulence in the thermodynamic limit.

\subsection{Comparison between full and approximated distributions}

 In the previous paragraphs it was found that in the thermodynamics limit, the distribution for the velocity can be approximated in the small velocity limit and for the tail by
\begin{equation}
P^{(\alpha)}(\bm{V})\sim
\begin{cases}
\frac{\left( c^{(\alpha)} \right)^{\alpha-3}\Gamma(4-\alpha)}{4\pi}  
 \exp\left(-\frac{1}{4  (c^{(\alpha)})^{3-\alpha}}\frac{\Gamma(6-2\alpha)}{\Gamma(3-\alpha)}V^2\right),& V<V_{gcut}\\\\
\frac{n \left(\gamma\lvert\Psi(\alpha)\rvert\right)^{\frac{2}{3-\alpha}}} {3-\alpha}V^{-2\frac{4-\alpha}{3-\alpha}},  &V> V_{cut},
\end{cases}
\label{totpdf1}
\end{equation}
for $\alpha\in(0,\,\,2)$, $c^{(\alpha)}$ defined as in \eqref{ca}, $\Psi(\alpha)$ defined as in \eqref{psi}, $V_{gcut}$ and $V_{cut}$ as in \eqref{Vgcut} and \eqref{Vcut} respectively. Both the variance and the slope of the tail depend thus on the degree of locality $\alpha$.

\begin{figure}
\begin{center}
\subfigure[]{
{\includegraphics[width=.5\textwidth]{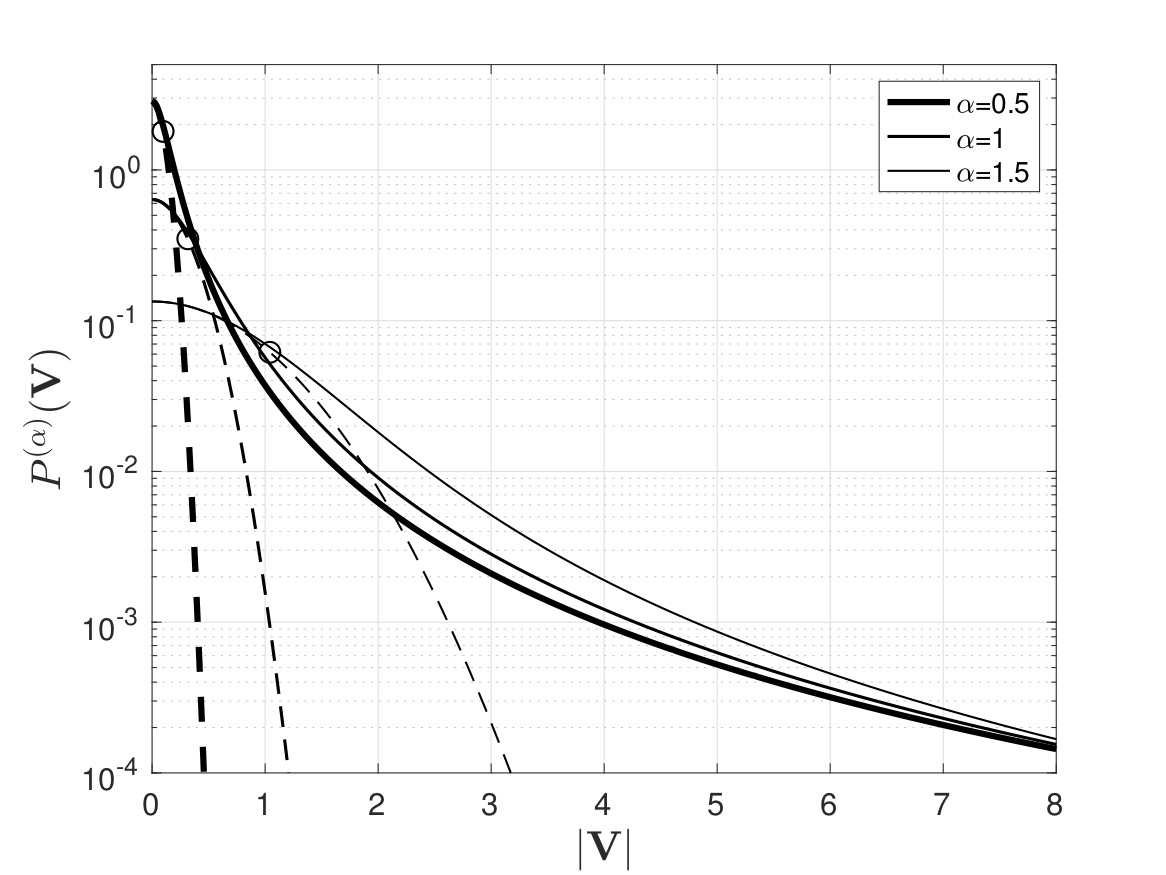}}}%
\subfigure[]{
{\includegraphics[width=.5\textwidth]{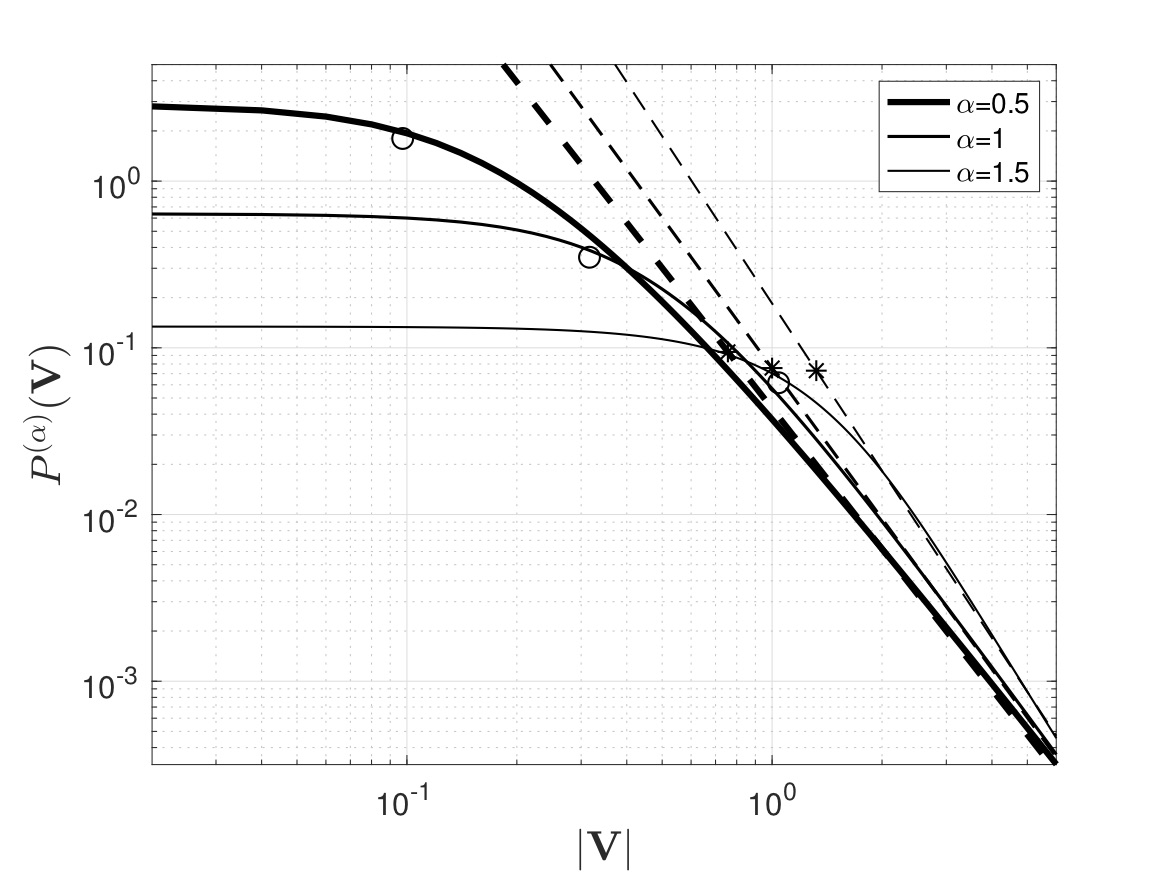}}}\\%
\subfigure[]{
{\includegraphics[width=.5\textwidth]{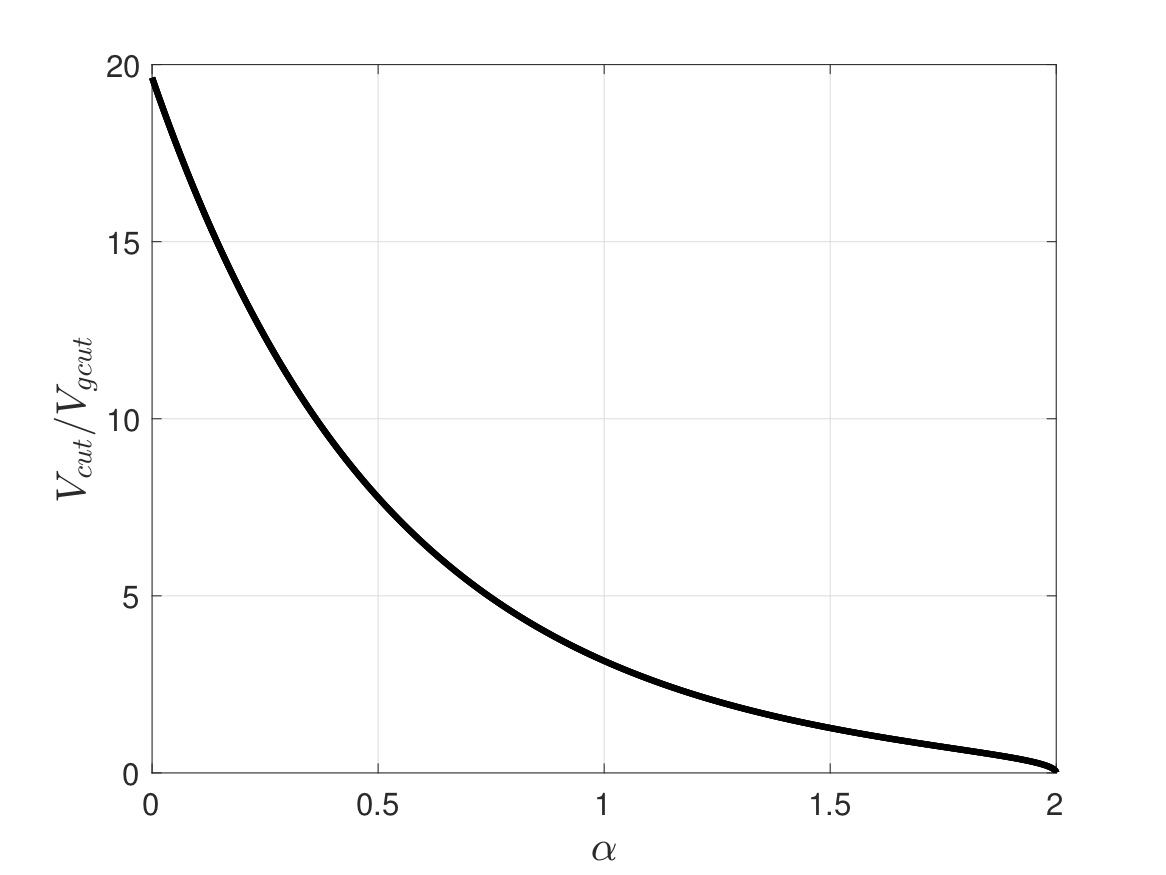}}}%
\subfigure[]{
{\includegraphics[width=.5\textwidth]{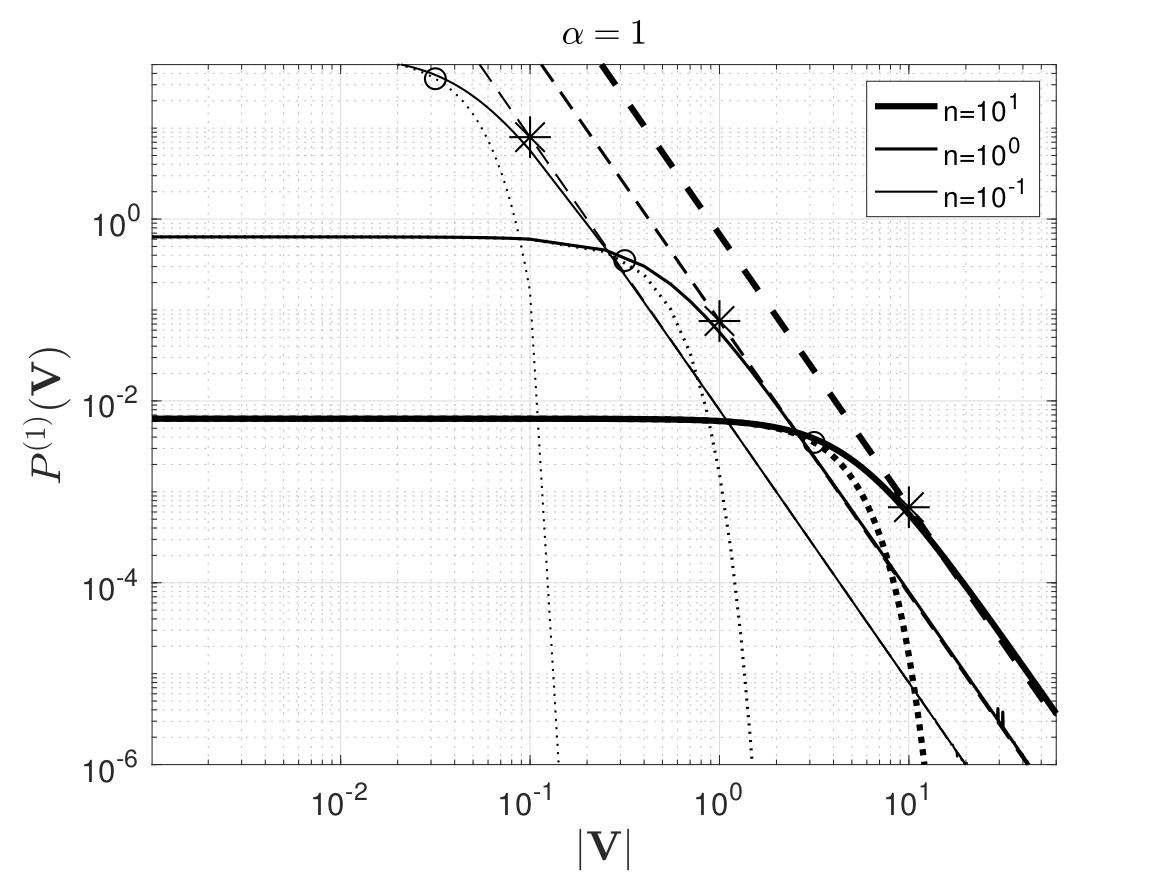}}}%
\caption{Comparison of the full distribution \eqref{pfull} with the approximation \eqref{totpdf1}. Panel (a) shows the PDFs (continuous lines) in semilogarithmic scale for fixed density $n=1$ and circulation $\gamma=1$, for different values of locality degree $\alpha=0.5, \,1,\,1.5$, and the corresponding Gaussian functions (dashed lines) for the small velocity limit approximation. The circles represent the respective values for $V_{gcut}$.   Panel (b) shows the previous distributions in logarithmic scale and the tail (dashed lines) approximations. The stars represent the respective values of $V_{cut}$. Panel (c) shows the ratio between  $V_{cut}$ and $V_{gcut}$. 
Panel (d) shows the case $\alpha=1$, SQG, for three different values of the density $n=10^{1},\,10^{0},\,10^{-1}$ and $\gamma=1$. As before dotted lines represent the small velocity approximations, dashed lines the power law tail approximations, and the circles and stars the limits of validity for the approximations. }%
\label{fig:full}
\end{center}
\end{figure}

Figure \ref{fig:full} shows a comparison between the full distribution obtained upon the integration of \eqref{pfull} and approximation \eqref{totpdf1}.

Figure \ref{fig:full}a shows the full PDFs (full lines) and the PDFs obtained from the small velocity limit approximation (dashed lines) for $\alpha= 0.5,\,1,\,1.5$.  For fixed values of the density $n$ and circulation $\gamma$, the distributions tends to be broader for increasing values of $\alpha$. This is better seen in 
Figure \ref{fig:full}b, which shows the distributions in logarithmic scale in order to highlight the behaviour of the tails and their  power law approximations. For increasing values of $\alpha$, the tails tend to become less important. This can be better understood by considering
the ratio between the limits of validity of the two approximations $V_{cut}$ and $V_{gcut}$, which is shown Figure \ref{fig:full}c.
It should be noted that this ratio is independent on the values of the density and circulation. 
When $\alpha \to 0$ the full distribution becomes extremely peaked and, although the limits of validity for both approximations decrease, the ratio between them has a maximum.  For increasing values of $\alpha$ the ratio decreases, reaching the value of zero for $\alpha\to2$, showing thus that the distribution tends to become a Gaussian with diverging variance. This is a consequence of the fact that in $2D$ turbulence the thermodynamic limit is not defined.
The ratio $V_{cut} / V_{gcut}\sim1$ when $\alpha\sim1.62$. For this value of $\alpha$ it is thus possible to build the full distribution simply by pasting  the two approximations together.

Finally, Figure \ref{fig:full}d  shows the SQG distribution, $\alpha=1$, and its approximations for different values of the density. In this case $V_{cut} / V_{gcut}\sim3.16$, so for the SQG  case the value of $V_{cut}$ is always around three times the value of $V_{gcut}$. 

\subsection{\label{sec:rem} Further analytical remarks}

\subsubsection{Non-neutral systems}
The distribution \eqref{totpdf1} has been derived by using assumption (i), i.e. assuming that the total system is neutral. If we want to take into account also a possible solid rotation for the system, we must consider that the average velocity increases linearly with the distance and the probability distribution gains a spatial dependency. At a point $\bm{r}\neq0$ the distribution \eqref{totpdf1} must be modified by replacing the velocity $\bm{V}$ with the fluctuation 
\begin{equation}
\bm{\mathcal{V}}=\bm{V}-\langle\bm{V}\rangle=\bm{V}-\frac{1}{2}n\gamma\bm{r}_{\perp},
\end{equation}
so that
\begin{equation}
P^{(\alpha)}(\bm{\mathcal{V}})\sim
\begin{cases}
\frac{\left( c^{(\alpha)} \right)^{\alpha-3}\Gamma(4-\alpha)}{4\pi}  
 \exp\left(-\frac{1}{4  (c^{(\alpha)})^{3-\alpha}}\frac{\Gamma(6-2\alpha)}{\Gamma(3-\alpha)}\mathcal{V}^2\right),& \mathcal{V}<\mathcal{V}_{gcut}\\\\

\frac{n \left(\gamma\lvert\Psi(\alpha)\rvert\right)^{\frac{2}{3-\alpha}}} {3-\alpha}\mathcal{V}^{-2\frac{4-\alpha}{3-\alpha}},  &\mathcal{V}> \mathcal{V}_{cut}.
\end{cases}
\label{totpdf}
\end{equation}

\subsubsection{Remarks on $\alpha \ge 2$}

In order to obtain the probability distribution, we have considered  $n C^{(\alpha)}(\bm{\rho})/N \ll1$ in the limiting process \eqref{ANL}. This limit can however be considered valid just for $\alpha \neq 2$. In the case in which $\alpha = 2$, Figure \ref{fig:kappa} shows in fact that $\kappa^{(\alpha)}$ diverges and the special limit can not be used. We can thus conclude that, although we could expect the variance to diverge when the dynamics becomes less local, the shape of the probability distribution is not to be considered correct when $\alpha\to 2$.

For $\alpha>2$, $\kappa^{(\alpha)}$ has a discontinuous behaviour, and assumes both positive and negative values. However, when $\kappa^{(\alpha)}>0$, e.g. in the interval $\alpha \in (2.5,\,\,2.66]$, one can use exactly the same strategy for the computation of the probability, both for the core and the tails. The results here derived will thus hold also for some of the $\alpha$-models in the nonlocal dynamics.

\subsubsection{\label{sec:levelequiv} Equivalence between nonuniform configurations of point vortices}

 In general for different values of $\alpha$ we can found different PDFs for the velocity field generated by the vortices. However, it is possible to connect the statistics of the velocity field of point vortices uniformly distributed for local $\alpha$-models in classical turbulence, with the statistics of the velocity field for point vortices non-uniformly distributed, that are relevant e.g. for quantum turbulence \citep{Bradley2012,Reeves2013,Reeves2014,Skaugen2016}. For example, using the damped Gross-Pitaevskii equation with a stirring potential, it is possible to simulate a statistically steady state turbulent regime of a $2$D trapped Bose-Einstein condensates, where  vortices are emitted in clusters in the wake of the stirring obstacle \citep{Skaugen2016-2}.  The statistic of the velocity field is an indicator of  the vortex clustering and can be used to determine if  the quantum turbulence exhibits an inverse energy cascade.

Consider an ensemble of $N$ identical point vortices on a disk of radius $R$ for an $\alpha$-model described by $\alpha_1=2$, and distributed following the probability density 
\begin{equation}
p_{(\beta)}(\bm{r})=\frac{n_{(\beta)}}{N}\lvert\bm{r}\rvert^{-\beta-1}, \quad \beta\in(-1,\,\,1),
\label{pfrac}
\end{equation}
where $\lvert\bm{r}\rvert$ is the distance to the origin of the disk and
\begin{equation}
n_{(\beta)}=\frac{N(1-\beta)}{2\pi R^{1-\beta}}
\label{densf}
\end{equation}
is the fractal density, that is, the  cluster is self-similar with fractal dimension $1-\beta$.  For $\beta=1/3$ the distribution of the clustered vortices is associated with an inverse cascade in $2$D quantum turbulence. Following \citet{Skaugen2016,Chavanis2009}, it is possible to find the probability for the vortices fluctuation as
\begin{equation}
P_{(\beta)}^{(2)}(\bm{V})=\frac{1}{(2\pi)^2}\int_0^{2\pi}\int_{0}^{\infty}\rho e^{-n_{(\beta)} C_{(\beta)}^{(2)}(\bm{\rho})}e^{i\rho V\cos\theta}\,d\rho\, d\theta,
\label{Pb}
\end{equation}
where
\begin{align}
C_{(\beta)}^{(2)}(\bm{\rho})&=2\pi\left(\frac{\gamma\rho}{2\pi}\right)^{1-\beta}\int_{0}^{\infty} (1-J_0(x)) x^{\beta -2}\,dx \nonumber \\
                            &=2\pi\, \kappa^{(2)}_{(\beta)}\,\left(\frac{\gamma\rho}{2\pi}\right)^{1-\beta},
\label{Cb}
\end{align}
being $\kappa^{(2)}_{(\beta)}$ the number coming from the integral, that for the values of $\beta$ in the interval considered is positive.

These results are similar to the ones found using uniform distribution for the point vortices and different $\alpha$-models  in the thermodynamic limit, that is $R,N\to \infty$ with constant density. The distribution of the velocity  for $\alpha_1=2$ point vortices non-uniformly distributed  with fractal dimension $1-\beta$, \eqref{Pb},   can be related to the 
distribution of the velocity for any $\alpha_2\in(0,\,\,2)$ point vortices distributed  uniformly in space, for classical turbulence, by setting
\begin{equation}
\beta=\frac{1-\alpha_2}{3-\alpha_2}\quad \forall \alpha_2 \in(0,\,\,2) \,\,\text{and}\,\,\alpha_1=2,
\label{beta}
\end{equation}
and rescaling the circulation of the $\alpha_1$-model as
\begin{equation}
\gamma\to\gamma\frac{2\pi\lvert\Psi(\alpha_2)\lvert}{(3-\alpha_2)^{\frac{3-\alpha_2}{2}}}.
\label{scale}
\end{equation}
Clearly, when the solution of \eqref{formalsol} is considered, we have also to substitute  $n_{(\beta)}$ with $n$ of \eqref{dens}.
Since we are considering local dynamics, i.e. $\alpha_2\in(0,\,\,2)$, then the only possible values for $\beta$ are in the interval $(-1,\,\,1/3)$. Note that $\beta=1/3$ can be considered only if $\alpha_2=0$, a non relevant case for classical turbulence.

The relations above show that the statistics of the  velocity  for  point vortices with circulation $\gamma$ uniformly distributed in  the SQG framework, $\alpha_2=1$, 
 are the same of the ones for point vortices with $\alpha_1=2$, but distributed as 
\begin{equation}
p_{(0)}(\bm{r})=\frac{n_{(0)}}{N}\lvert\bm{r}\rvert^{-1},
\label{pfrac0}
\end{equation}
where
\begin{equation}
n_{(0)}=\frac{N}{2\pi R},
\label{densf0}
\end{equation}
and with a circulation scaled following \eqref{scale}, that is $\gamma/\pi\lvert\Psi(1)\rvert$.

It is clearly also possible to make a generalization  relating the statistics of $\alpha_2$-model, characterized by uniform distribution of the vortices, with $\alpha_1$-model, $\alpha_1\neq2$, that are distributed following a  power law distribution using 
\begin{equation}
\beta=\frac{2\alpha_1-\alpha_2-3}{3-\alpha_2}\quad \forall \alpha_2 \,\,\text{and}\,\,\alpha_1\neq2~,
\end{equation}
and setting in the $\alpha_1$-model
\begin{align}
\gamma&\to\frac{\gamma}{\Psi(\alpha_1)}\left(\frac{3-\alpha_1}{4\pi^2}\right)^{3-\alpha_2} \quad  \alpha_2=2 \,\,\text{and}\,\,\alpha_1\neq2~,
\end{align}
and
\begin{align}
\gamma&\to\gamma\left(\frac{3-\alpha_1}{3-\alpha_2}\right)^{\frac{3-\alpha_2}{2}}\quad  \alpha_2\neq2 \,\,\text{and}\,\,\alpha_1\neq2~.
\end{align}
For this general group of transformations $\beta\in(-5,\,\,1/3)$. 
Since we are studying a cluster of vortices in the plane, we could expect a fractal dimension  $(1-\beta)\in(0,\,\,2)$ and so,  only the combination of $\alpha_1$ and $\alpha_2$ for which $\beta\in(-1,\,\,1/3)$ should be considered.

\section{Numerical simulations and finite size effects\label{sec:Numerical Simulations}} 

The analytical results illustrated  in this work make use of the point vortex approximation, a strong simplification of the original problem. In order to test the reliability of the results when a fully turbulent field is considered without approximations, we perform numerical simulations
of \eqref{geul} and \eqref{coupl} for three different values of $\alpha$. In particular we consider $\alpha=1$, which correspond to SQG, and other two turbulent fields with degrees of locality that are smaller and greater than the SQG case, that is $\alpha=1.5$, and $\alpha=0.5$ respectively. 

We will focus on the slope of the tails of the PDFs for the absolute value of the velocity field, which depends only on the degree of locality $\alpha$. In order to preserve the Hamiltonian structure of the system, the simulations are performed in a freely decaying setup.  
%

The advection of the $\zeta$ field is performed using a semi-spectral method in a double periodic domain of length $L=2\pi$. 
The horizontal resolution is set to $512\times512$ Fourier modes.  We use a fourth order Runge-Kutta time stepping scheme 
with time step $\Delta t=0.005$. The simulations are performed up to $T=800$. The Jacobian is computed 
 using the  \cite{ARAKAWA1966119} discretization. To avoid accumulation of enstrophy at the small scales, we employ an exponential filter multiplying each Fourier mode 
 by $\rho(2K_x/N)\rho(2K_y/N)$, where $\rho(x)=\exp(-a x^m)$ \citep{HOU2007379} and $K_x,~K_y$ indicate, respectively, the zonal and meridional wavenumbers. Following \citet{constantin2012new,ragone2016study} the coefficients of the filter are set to $a=36$ and $m=19$. This choice of the parameters  ensures that approximately two-third of the low wavenumber modes remains unchanged.

The initial condition is set by a random $\zeta$ field obtained by inversion of the stream function
\begin{equation}
\hat{\psi}(\textbf{k},0)\propto\frac{k^{m/4 -1}}{(k+k_0)^{m/2}},
\end{equation}
with $m=25$ and $k_0=6$ \citep{Hakim2002,ragone2016study} and where the hat indicates the Fourier transform of the quantity specified in physical space. 
The initial conditions for the three different simulations, corresponding to $\alpha=(1.5,\, 1,\, 0.5)$, have been normalized so that the quantity $E(\textbf{k})=1/2\, k^2 \lvert\hat{\psi}(\textbf{k},0)\lvert$, that in the SQG case represents the surface kinetic energy, is $E(\textbf{k})=4\pi^2(3,\,3,\,80)$ respectively. The higher value for the case of $\alpha=0.5$ is given by the fact that in this case the PDF of the velocity tends to be more peaked and the determination of the tail can thus be problematic. A more energetic field ensures thus a larger number of points in the tails of the distribution.

Each simulation reaches a quasi-steady state for which the two conserved quantities of the model, that is the generalized energy $E_g$ and the generalized enstrophy $Z_g$, defined as
\begin{equation}
E_g=-\overline{\psi\zeta}, \qquad  Z_g=\frac{1}{2}\overline{\zeta^2},
\end{equation}
where the overbars indicate domain averages, are approximately constant.
Once this quasi-steady state is achieved, we compute  $P(\lvert\textbf{V}\rvert)$ for several eddy turnover times, defined for different values of the degree of locality $\alpha$ as
\begin{equation}
T_{eddy} = \frac{(2\pi)^{2-\alpha}}{\zeta_{rms}}.
\label{ted}
\end{equation}
The PDFs are then averaged in time between $T=400$ and $T=800$.
In the simulations here performed one has  $T_{eddy} \sim6.12$ for $\alpha=1.5$ and $T_{eddy} \sim 36.5$ for $\alpha=1$ and $\alpha=0.5$. The time average for the PDFs correspond thus averages over 65 eddy turnover times for $\alpha=1.5$ and 11 eddy turnover times for $\alpha=1$ and $\alpha=0.5$.

In the point vortex approximation, the velocity field decrease with power law \eqref{vr}, which has a maximum at the location of the vortex. Considering vortices with finite size,  the velocity field grows from the center of the vortices, reaching a maximum at the vortex boundary, and decreases outside. To recover a situation similar to the one obtained using the point vortex approximation, we need to use a mask that allow us to consider the velocity field in between the vortices, excluding their interiors. This is approximately  achieved considering region of the velocity field for which $\zeta\in[-\sigma(\zeta),\,\sigma(\zeta)]$, where $\sigma(\zeta)$ is the standard deviation of the $\zeta$ field. For the case $\alpha=0.5$ the strong forward cascade of $\zeta$ variance produces a large number of small vortices. In this case we thus use the interval $\zeta\in[-\sigma(\zeta)/2,\,\sigma(\zeta)/2]$ to avoid their inclusion in the region used for the computation of the PDFs. 


\subsection{SQG ($\alpha = 1$)}

\begin{figure}
\begin{center}
\subfigure[$\zeta$]{
{\includegraphics[width=.5\textwidth]{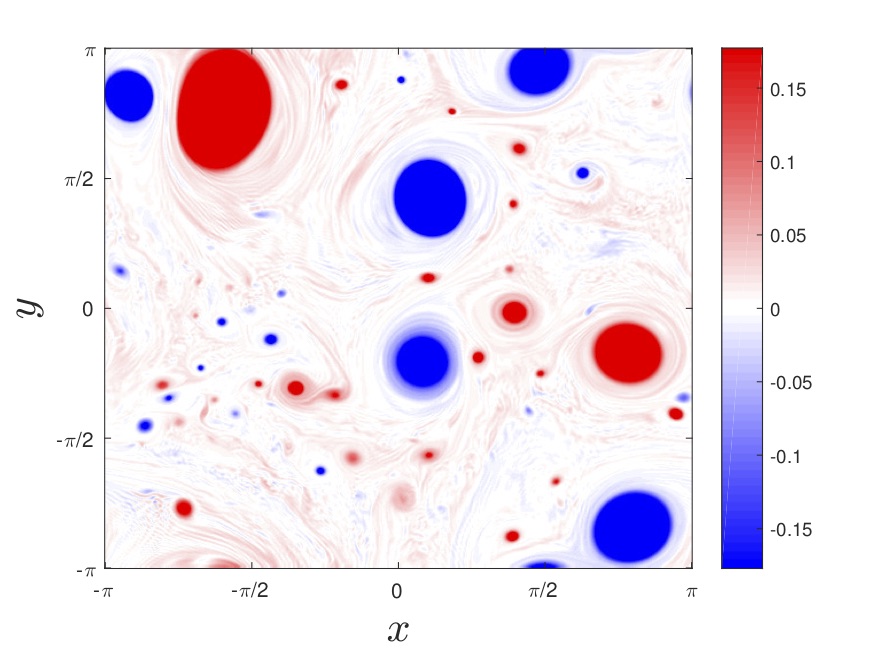}}}%
\subfigure[Mask]{
{\includegraphics[width=.5\textwidth]{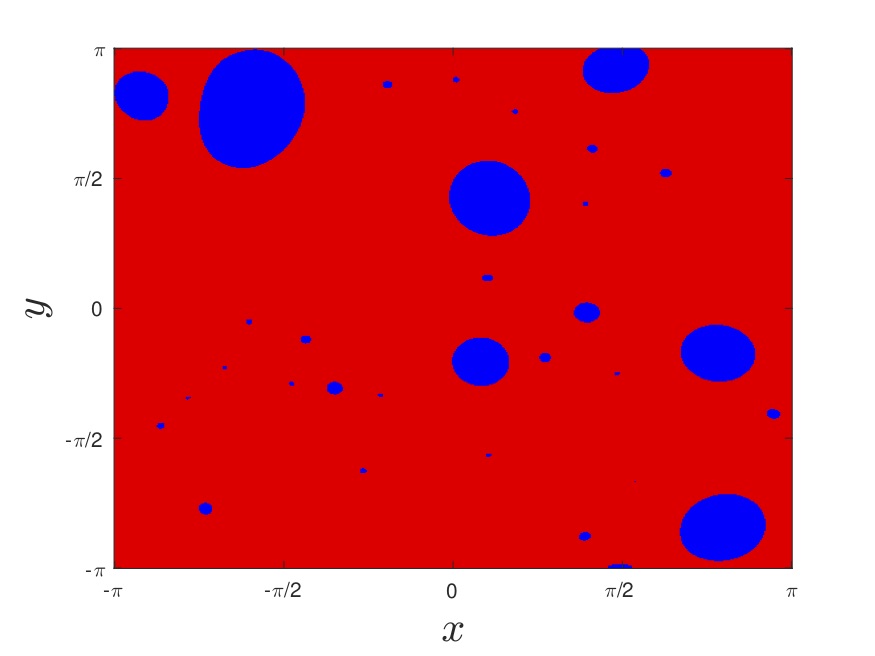}}}\\%
\subfigure[$\lvert\textbf{V}\rvert$]{
{\includegraphics[width=.5\textwidth]{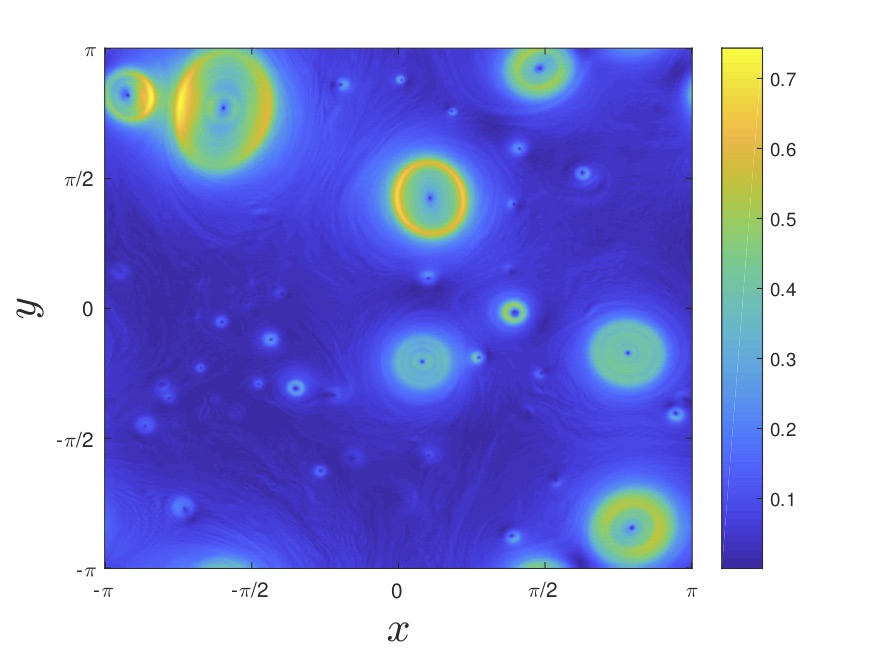}}}%
\subfigure[$P(\lvert\textbf{V}\rvert)$]{
{\includegraphics[width=.5\textwidth]{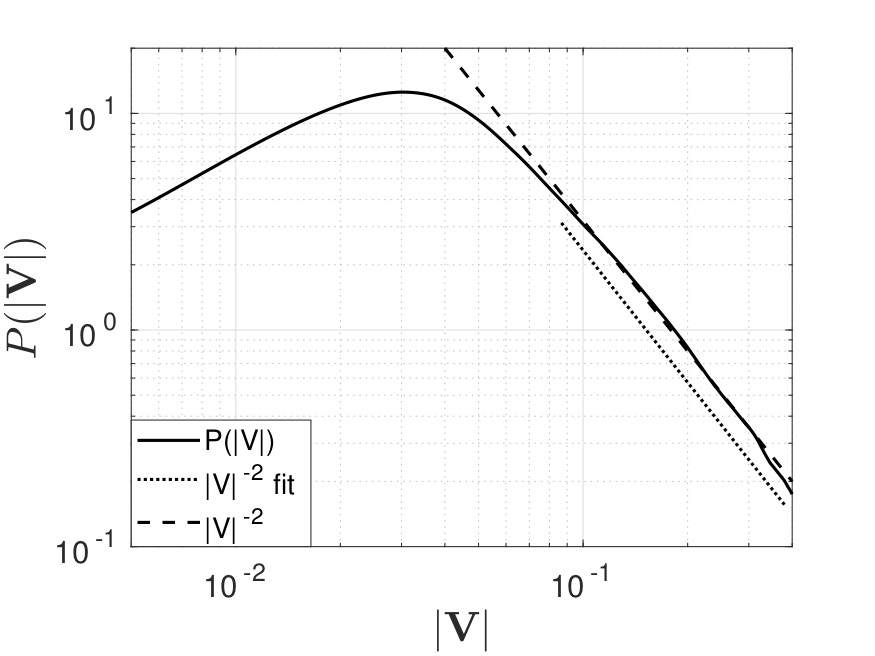}}}%
\caption{Numerical simulation of \eqref{geul} and \eqref{coupl} with $\alpha=1$. Panel (a) shows the last snapshot of the simulation for the $\zeta$ field.  The range of variability is limited to  $[-\sigma(\zeta),\,\sigma(\zeta)]$. Panel (b) shows the mask used for the computation of the PDF with respect to the previous panel. 
Panel (c) shows the map for the module of the velocity, $\lvert\textbf{V}\rvert$, in the last snapshot of the simulation. Panel (d) shows the PDF (solid line) for the module of the velocity computed over several eddy turnover time \eqref{ted}, the power law approximation \eqref{tailmod} (dashed line), and the fit of the tail of the PDF (dotted line). 
}%
\label{fig:n1}
\end{center}
\end{figure}

Figure \ref{fig:n1} shows the SQG case, which corresponds to $\alpha=1$. The $\zeta$ field (Figure \ref{fig:n1}a) shows several well defined vortices as well as a number of  secondary, smaller vortices in between them.  Panels (b) and (c) show that the mask is able to discriminate between the regions inside and outside the vortices. The velocity field shows that inside the vortices the velocity increases from the center to the boundaries and decreases quickly outside of them (Figure \ref{fig:n1}c).

For the SQG case, \eqref{tail} yields
\begin{align}
P^{(1)}(\lvert\bm{V}\lvert=V)\sim\frac{2\pi n\gamma\lvert\Psi(1)\rvert}{2}V^{-2}~.
\end{align}
%
The averaged PDF (Figure \ref{fig:n1}d) shows indeed tails with slope $-2$, which is in agreement with the value predicted by the analytical point vortex model. 

\subsection{$\alpha = 1.5$}

\begin{figure}
\begin{center}
\subfigure[$\zeta$]{
{\includegraphics[width=.5\textwidth]{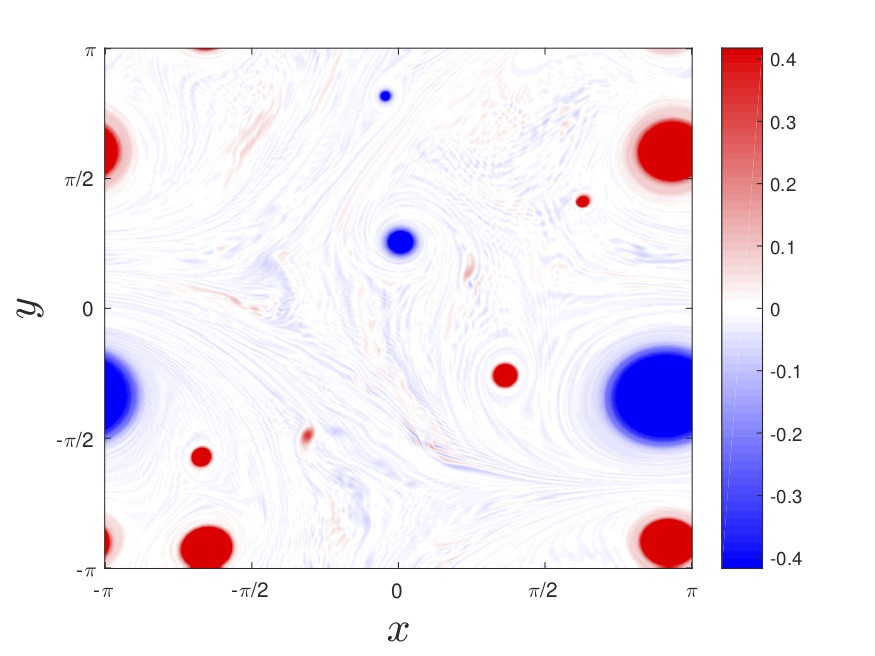}}}%
\subfigure[Mask]{
{\includegraphics[width=.5\textwidth]{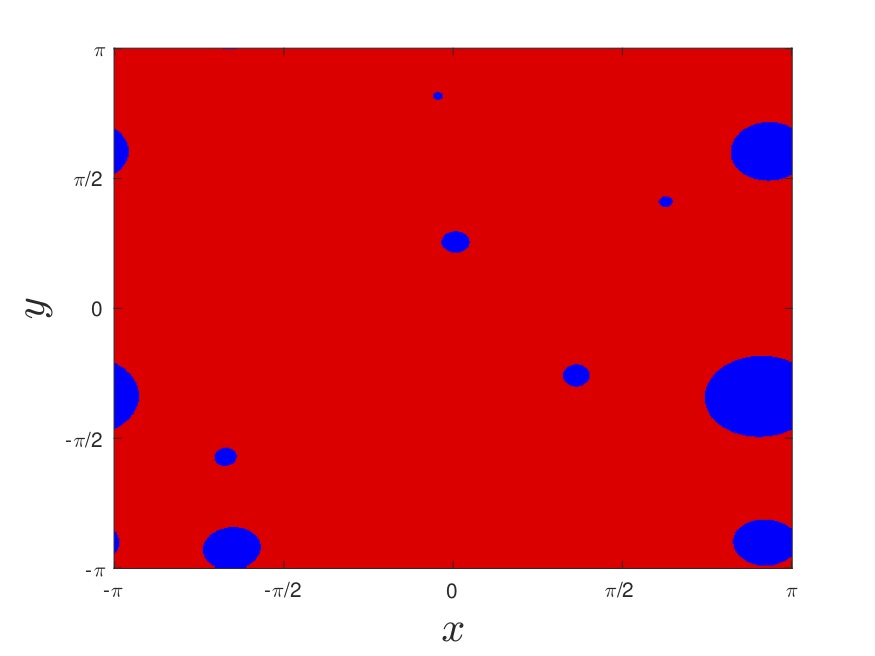}}}\\%
\subfigure[$\lvert\textbf{V}\rvert$]{
{\includegraphics[width=.5\textwidth]{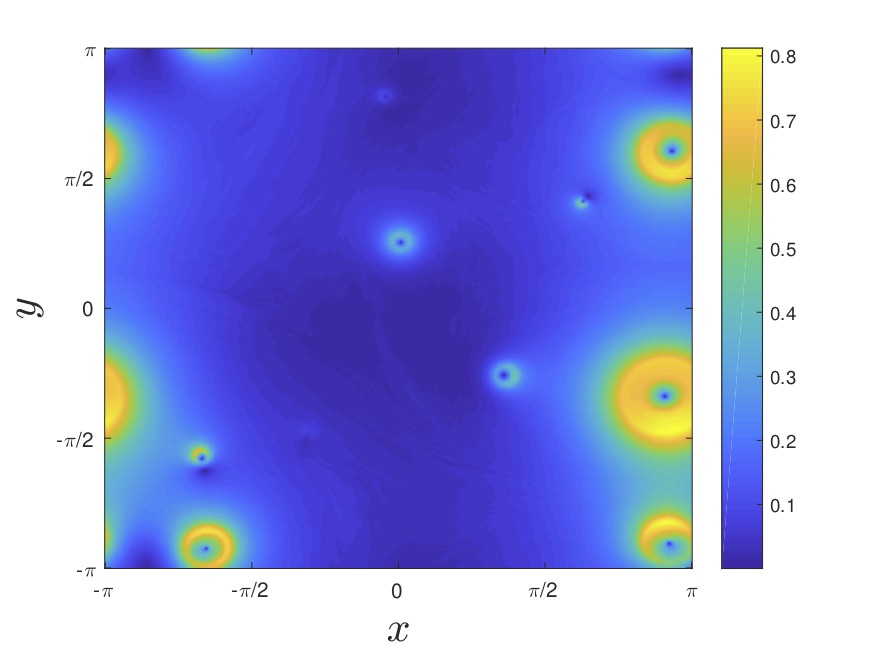}}}%
\subfigure[$P(\lvert\textbf{V}\rvert)$]{
{\includegraphics[width=.5\textwidth]{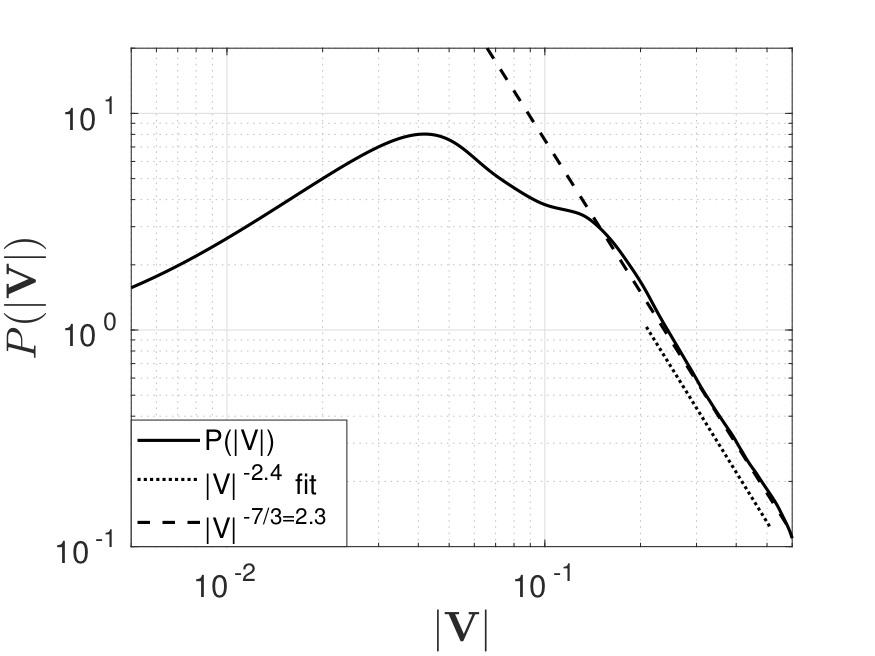}}}%
\caption{As in Figure \ref{fig:n1} but for $\alpha=1.5$. 
 }%
\label{fig:n1.5}
\end{center}
\end{figure}

Figure \ref{fig:n1.5} shows the results for the case $\alpha=1.5$. The $\zeta$ field (Figure \ref{fig:n1.5}a) shows several well defined vortices with a smaller number of  secondary vortices in between them than for SQG.  Also in this case, panels (b) and (c) show that the mask is able to discriminate between the regions inside and outside the vortices. 
The comparison between Figures \ref{fig:n1.5}c and \ref{fig:n1}c shows that for $\alpha=1.5$ the vortices have broader borders, characterised by high velocities, than for SQG.  
For $\alpha=1.5$, the analytical results predict a slope for the tails of the PDF of $-7/3=-2.\bar{3}$. The numerical results show that the tail of the averaged PDF for the absolute value of the velocity field follows almost exactly the power law found by the analytical result,  with a fit for the tail region highlighted by the length of the dotted line of $-2.4$ (Figure  \ref{fig:n1.5}d). Notice that the core of the PDF exhibits a secondary peak caused by the finite size of the vortices and not predicted by the point vortex approximation. 

\subsection{$\alpha = 0.5$}

\begin{figure}
\begin{center}
\subfigure[$\zeta$]{
{\includegraphics[width=.5\textwidth]{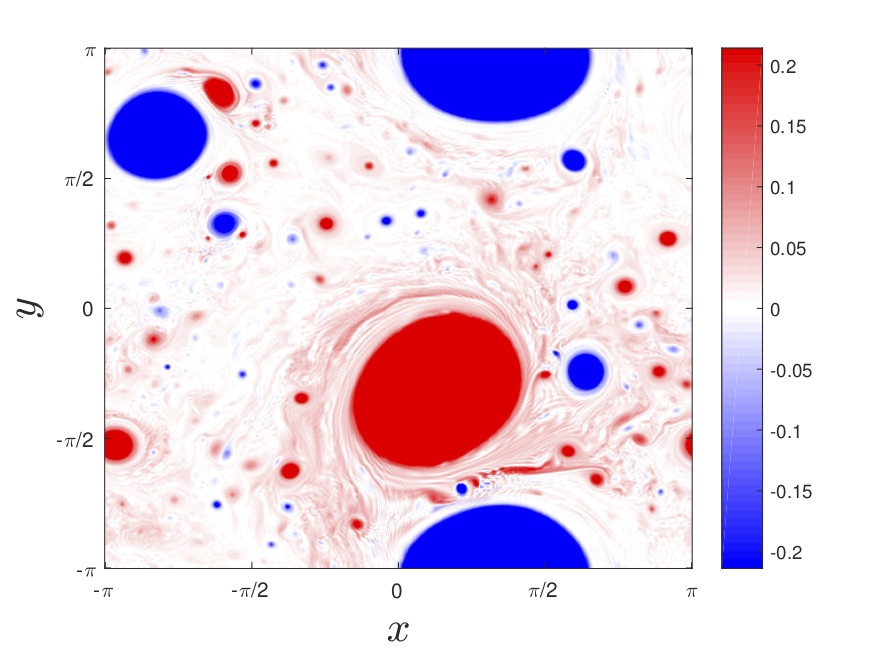}}}%
\subfigure[Mask]{
{\includegraphics[width=.5\textwidth]{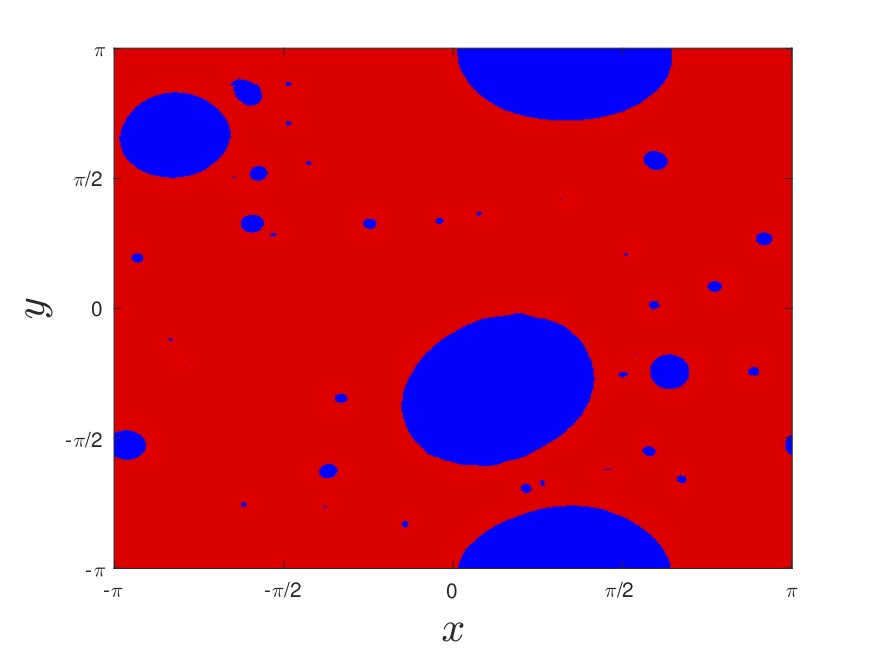}}}\\%
\subfigure[$\lvert\textbf{V}\rvert$]{
{\includegraphics[width=.5\textwidth]{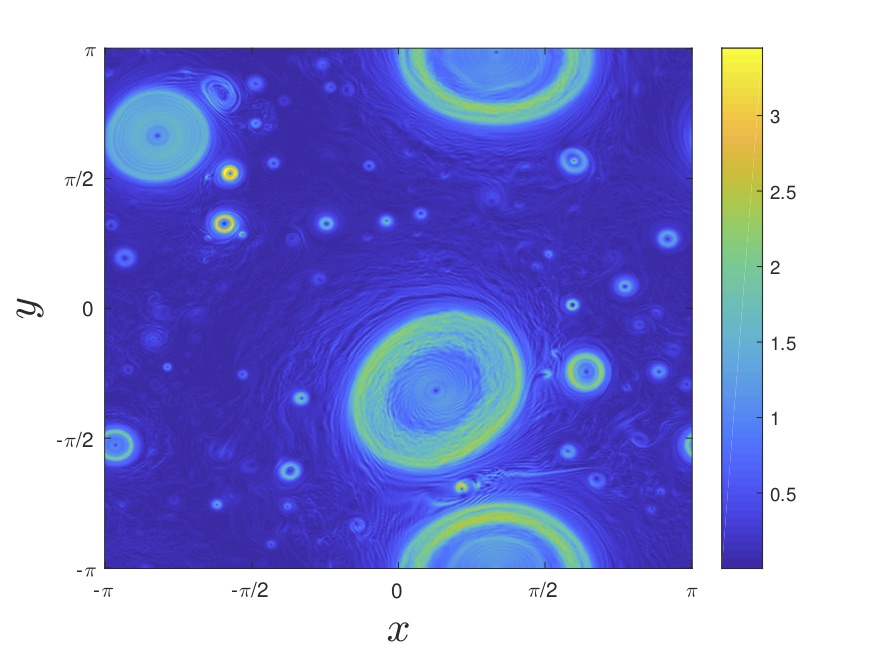}}}%
\subfigure[$P(\lvert\textbf{V}\rvert)$]{
{\includegraphics[width=.5\textwidth]{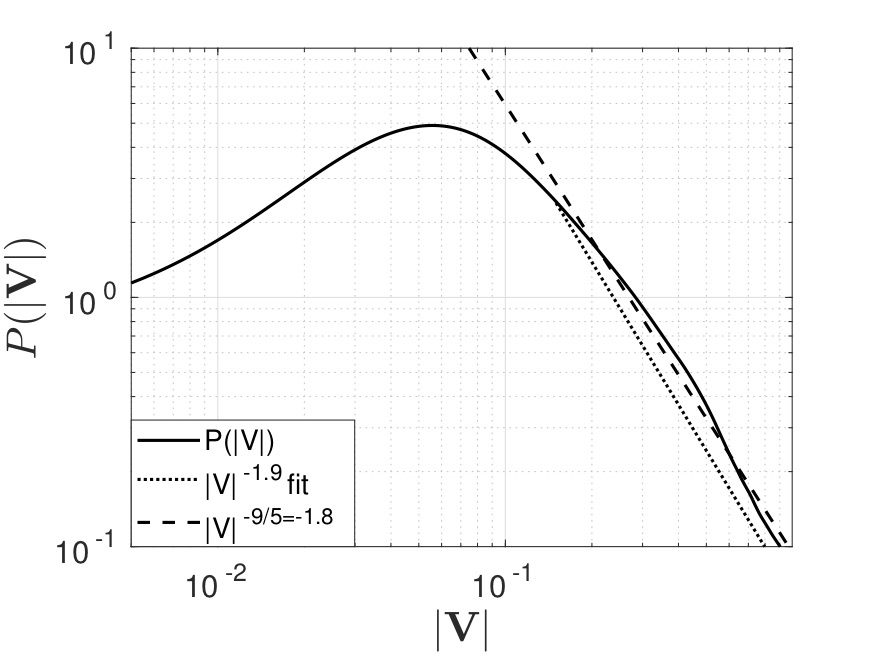}}}%
\caption{As in Figure \ref{fig:n1.5} but for $\alpha=0.5$. In this case the range of variability of $\zeta$, and the corresponding mask, are in the interval $[-\sigma(\zeta)/2,\,\sigma(\zeta)/2]$}%
\label{fig:n0.5}
\end{center}
\end{figure}

Finally, Figure  \ref{fig:n0.5} shows the results for $\alpha=0.5$. The $\zeta$ field (Figure  \ref{fig:n0.5}a) shows strong secondary instabilities which
span big portion of the space  in the domain.
The numerically estimated slope of the tails for this case is of $-1.9$, which is in good agreement with the value predicted by the theory of $-9/5=-1.8$. 


Notice that the finite size effects of the vortices could be introduced in the analytical model by the inclusion of
a finite radius $a$, so the maximum allowable velocity is $V=\gamma \lvert\Psi(\alpha)\lvert/(2 a)^{3-\alpha}$ obtained when the two vortices are at distance $\approx 2 a$ from each others. Infinite velocity and collapse of vortices are avoided. On the phenomena of collapse, see e.g. \citet{aref1979motion,novikov1979vortex,o1987stationary,tavantzis1988dynamics,kimura1990parametric,leoncini2000motion,leoncini2001chaotic,hernandez2007collisions,sakajo2008non,sakajo2012instantaneous,kudela2014collapse,BadinBarry2018}.

\section{\label{sec:conclusion} Conclusions }
In this work we have investigated the velocity statistics for turbulent flows, with the simplification of point vortices randomly distributed. The interactions of the point vortices are ruled by a family of $\alpha$-models in the local  dynamics, which include the physically and mathematically interesting SQG model.  
The local dynamics differ from the case of $\alpha=2$ as we are able to consider the true thermodynamic limit, that is, if $N$ is the number of point vortices into a disc of radius $R$, we can consider the limits $N\to\infty$, $R\to\infty$ keeping the density $n$ constant. This limit is not defined for $\alpha=2$. As a consequence of this fact, the distributions that we obtain are well defined and independent on the number of point vortices in the domain considered.

This result clearly shows that the case $\alpha=2$ must be considered as a singular limit for the dynamics. The passage from local to nonlocal dynamics corresponds to a change in the topology of the system, and the passage between the two regimes must not be considered as continuous.

Results show that he central region of the distribution is not a Gaussian curve, in contrast to the case of $2D$ turbulence. However, in the small velocity limit we can use  a Gaussian function to approximate the peak of the distribution. 
The tail of the distribution follows a power law.
Both the variance of the Gaussian approximation and the slope of the tail depend parametrically on $\alpha$. 
 It is interesting to remark that the tail of the distribution exhibits self-similarity in respect to the density parameter.  

Turbulence in the atmosphere and in the ocean is often characterized by means of wavenumber turbulent spectra. These can be used to infer for example if dynamics are either local or nonlocal. Very often the slopes of the inertial range in geophysical fluids are however difficult to obtain from observations or they can be ambiguous in their interpretation. The results obtained in  the current study suggest however that the local nature of the turbulent fluctuations might be deduced from the shape of the tails of the PDFs. The velocity's PDFs  for the barotropic turbulence in the ocean has been studied in \citet{Bracco2000,Bracco2000_2}


Finally, we have also observed a connection between  the statistics of the velocity field of point vortices uniformly distributed for local $\alpha$-models in classical turbulence, with the statistics of the velocity field for point vortices non-uniformly distributed, that are relevant  for quantum turbulence. 
In particular the statistics of the  velocity  for  point vortices uniformly distributed in  the SQG framework, are the same of the ones for point vortices following the $2$D quantum turbulence and distributed following a probability density $p(\bm{r})\sim\bm{r}^{-1}$. 

It should be noted that the $\alpha$-models here analysed can be used to represent also higher order balanced models of geophysical flows, such as the surface semi-geostrophic (SSG) model \citep{badin2013surface,ragone2016study}. The velocity fluctuations for these higher order models are still unexplored and might give important insights on the physics of these systems.

\bigskip

\noindent
{\bf Acknowledgements}
\medskip

\noindent
GB would like to thank A.M. Barry for discussions on point-vortex dynamics in the $\alpha$-models.
This study was partially funded by the DFG research grant BA 5068/9-1.
GB was partially funded also by the DFG research grants TRR181, BA 5068/8-1 and 1740.

\appendix
\section{Derivation of \eqref{tail}}
Following \cite{Chavanis1999,Chavanis2009} and setting $t=\cos\theta$ and $z=\rho V$, \eqref{formalsol} can be written as
\begin{equation}
P^{(\alpha)}(\bm{V})=\frac{1}{2\pi^2V^2}\int_{-1}^1\frac{dt}{\sqrt{1-t^2}}\int_0^{\infty} ze^{izt} e^{-n C^{(\alpha)}(z/V)}dz,
\label{ptrans}
\end{equation}
where we have exploited the symmetry of the cosine function  to restrict the polar integration between $0$ and $\pi$. These substitutions allow to study the velocity PDF when $V\to\infty$ as a power series of $z/V$. In order to perform the expansion and integrate term by term, it is necessary to consider the variables as complex, and change the path of the integration to ensure the analyticity of the inner integrand. The path of $t$ can be deformed to the unit semicircle $\tau$ in the positive imaginary half plane. In this way $\arg t$ can vary continuously between $0$ and $\pi$. Rotating the integration path for $z$ of an angle $\omega(t)$ that depends on $\arg t$, we can ensure the convergence for $e^{izt}$. The real part of $izt$ must be negative, that is $\arg(izt)=\pi/2 +\omega(t)+\arg t$ must be chosen so that it is kept between $\pi/2$ and $3\pi/2$.  In the same way the real part of $C^{(\alpha)}(z/V)\propto z^{2/(3-\alpha)}$ should be positive, so that  also the second exponential goes to zero for large $z$. This means that  $\arg(z^{2/(3-\alpha)})=(2/(3-\alpha))\omega(t)$ should be kept in the interval $-\pi/2$ and $\pi/2$. In order to satisfy these constraints a possible choice for $\omega$ is
\begin{equation}
\omega(t)=\frac{1}{8}\left(\frac{\pi}{2}-\arg t\right).
\end{equation}
In this way $\omega(t)$ varies between $-\pi/16$ and $\pi/16$ and then $(2/(3-\alpha))\omega(t)$ respects the right condition explained above. After these necessary prescriptions we can expand the integrand
\begin{align}
P^{(\alpha)}(\bm{V})&=\frac{1}{2\pi^2V^2}\sum_{m=0}^{\infty}\frac{1}{m!}\left(-\frac{2\pi \kappa^{(\alpha)} n}{3-\alpha}\right)^{m}\left(\gamma \lvert\Psi(\alpha)\rvert\right)^{\frac{2m}{3-\alpha}}
\nonumber \\
&\times\left(\frac{1}{V}\right)^{\frac{2m}{3-\alpha}}
 \int_{\tau}\frac{dt}{\sqrt{1-t^2}}\int_{\omega(t)} ze^{izt} z^{\frac{2m}{3-\alpha}}dz~.
\end{align}
These integrals are convergent along any line where  the real part of $izt$ is negative, so we can rotate again the axis  for $z$ considering the rotation
$\omega(t)=\pi-\arg t$. Along this new integration path we also change the variable, $y=-izt$, obtaining
\begin{align}
P^{(\alpha)}(\bm{V})&=-\frac{1}{2\pi^2V^2}\sum_{m=0}^{\infty}\frac{1}{m!}\left(\frac{-2\pi \kappa^{(\alpha)n}}{3-\alpha} \right)^{m}\left(\gamma \lvert\Psi(\alpha)\rvert\right)^{\frac{2m}{3-\alpha}}
\nonumber \\
&\times\left(\frac{i}{V}\right)^{\frac{2m}{3-\alpha}}
 \int_{\tau}\frac{dt}{\sqrt{1-t^2}t^{2+\frac{2m}{3-\alpha}}}\int_0^{\infty} e^{-y} y^{\frac{2m}{3-\alpha}+1}dy~.
\end{align}
Since 
\begin{equation}
\int_{\tau}\frac{dt}{\sqrt{1-t^2}t^l}=-\frac{i}{2}(1-e^{i\pi l})B\left(\frac{m}{2},\frac{1}{2}\right),
\end{equation}
the first term of the expansion, $m=0$, vanishes (see, e.g. \cite{Skaugen2016}).
For $m=1$ we have 
\begin{align}
\int_{\tau}\frac{dt}{\sqrt{1-t^2}t^{2\frac{4-\alpha}{3-\alpha}}}
=-\frac{i}{2}(1-e^{-i\pi 2\frac{4-\alpha}{3-\alpha}})B\left(\frac{4-\alpha}{3-\alpha},\frac{1}{2}\right),
\end{align}
while the integral on $y$ for the $m=1$ is the $\Gamma$ function
\begin{equation}
\int_{0}^{\infty}e^{-y}y^{\frac{5-\alpha}{3-\alpha}}dy = \Gamma\left(2\frac{4-\alpha}{3-\alpha}\right).
\end{equation}
Combining all the results together \eqref{tail} is obtained at the leading  order.

\bigskip

\bibliographystyle{gGAF}
\bibliography{ms}
\vspace{12pt}

\end{document}